\documentclass[11 pt]{article}
\usepackage{amssymb,latexsym,amsmath,amsfonts}
\usepackage{graphicx, epsf}
\usepackage{epsfig}
\usepackage{subfigure}

\usepackage[active]{srcltx}

\def\ep{\varepsilon}

\def\kcb{\bar{k}_c}

\def\bfrho{\mbox{\boldmath$\rho$}}

\newtheorem{teo}{Theorem}

\newtheorem{propos}[teo]{Proposition}

\begin{document}

\title{Pattern formation driven by cross--diffusion\\in a 2D domain}

\author{G. Gambino\footnote{Department of Mathematics, University of Palermo, Italy, gaetana@math.unipa.it}$\;$
M.C. Lombardo\footnote{Department of Mathematics, University of Palermo, Italy, lombardo@math.unipa.it}$\;$
M. Sammartino\footnote{Department of Mathematics, University of Palermo, Italy, marco@math.unipa.it}}


\maketitle


\begin{abstract}
In this work we investigate the process of pattern formation in a two dimensional domain for
a reaction-diffusion system with nonlinear diffusion terms and the  competitive Lotka-Volterra
kinetics.
The linear stability analysis shows that cross-diffusion, through Turing bifurcation, is
the key mechanism for the formation of spatial patterns .
We show that the bifurcation can be regular, degenerate non-resonant and resonant.
We use multiple scales expansions to derive the amplitude
equations appropriate for each case and show that the system supports patterns like rolls,
squares, mixed-mode patterns, supersquares, hexagonal patterns.
\end{abstract}




\section{Introduction}\label{Sec1}

The aim of this paper is to study the pattern formation for the reaction-diffusion system:

\begin{subequations}\label{1.1}
\begin{equation}\label{1.1a}
\begin{split}
\frac{\partial u}{\partial t}=&\mathbf{\nabla} \cdot \mathbf{J}_1
+\Gamma u(\mu_1 -\gamma_{11}u -\gamma_{12}v),
\\\frac{\partial v}{\partial t}=&
\mathbf{\nabla} \cdot \mathbf{J}_2 +\Gamma v(\mu_2 -\gamma_{21}u
-\gamma_{22}v).
\end{split}
\end{equation}

\noindent where the fluxes $\mathbf{J}_i$ have the following nonlinear expressions:

\begin{equation}\label{1.1b}
\begin{split}
\mathbf{J}_1=&\mathbf{\nabla}(u (c_1+a_1 u +b v)),
\\\mathbf{J}_2=&\mathbf{\nabla}(v (c_2+a_2 v +b_2 u)).
\end{split}
\end{equation}
\end{subequations}

Here $u(\mathbf{x}, t)$ and $v(\mathbf{x}, t)$ are the population densities of two
competing species, and $\mathbf{x}\in \Omega$ with $\Omega=[0,L_x]\times[0,L_y]$.
The above system is supplemented with initial data and following Neumann boundary conditions:
$$
\mathbf{n}\cdot \mathbf{J}_1 =\mathbf{n}\cdot\mathbf{J}_2=0 \qquad \mbox{when} \quad \mathbf{x}\in \partial\Omega\; .
$$

The nonlinear diffusion terms describe the tendency of the species to diffuse, when densities
are high, faster than predicted by the usual linear diffusion towards lower density areas.
The parameters $a_i$ and $c_i$ are the self-diffusion and the linear
diffusion coefficients respectively, while the parameters $b$ and $b_2$ are the
cross-diffusion coefficients. For two competing species it is natural to suppose all these parameters
to be non negative.

The constants $\gamma_{ij}\geq 0$ are the competitive interaction coefficients,
the constants $\mu_i$ are the rates at which each species would grow in absence of competition, while
the parameter  $\Gamma$ regulates the size of the spatial domain
(or the relative strength of reaction terms).

Since the seminal paper of Turing \cite{Tur}, reaction-diffusion equations are one of the best-known
theoretical models  explaining self-regulated pattern formation in many different areas of physics,
chemistry, biology, geology, etc.
Turing showed that the interplay of diffusion and kinetics can destabilize the uniform steady state
and generate stable, stationary concentration patterns.
However, if the model has a trivializing kinetics, as it is the case for
competitive Lotka-Volterra kinetics, classical diffusion is not sufficient
to destabilize the equilibria, no matter what the diffusion rates are,
and no pattern formation can be observed. Thus, in order
to model segregation keeping a simple form for the kinetic term,
Shigesada, Kawasaki and Teramoto \cite{SKT79} proposed the
nonlinear evolution system \eqref{1.1}.

Strongly coupled parabolic systems with nonlinear
diffusion terms of the form given in \eqref{1.1} have been used to model different
physical phenomena and appeared in many context like chemotaxis \cite{LAK82,AF04}, ecology \cite{GHPSM07,TLP10,Jun10,ZLW11,GV11,Gal12,R-BT12},  social systems \cite{MK80,YPM04,Ep1}, turbulent transport in plasmas \cite{DCL02}, drift-diffusion in semiconductors \cite{CJ07,BDG96,DGJ97}, granular materials \cite{AT02,GJV03} and cell division in tumor growth \cite{Sherratt2000}.

The system \eqref{1.1} has been extensively investigated from the mathematical point of view.
In \cite{CJ04,CJ06,WF09}  global existence and regularity results have been obtained,
while in \cite{LouNi96} the existence of non constant steady state
solutions in the time independent case was investigated.
The proof of existence and stability of traveling wave solutions has
been obtained in \cite{WZ05}.
The existence of positive steady-state solutions in relation to
large cross-diffusion coefficients has been discussed in \cite{Ru96}.
Some families of exact solutions have been constructed in \cite{CM08} using the Lie symmetry approach while
Lyapunov functionals have been used in \cite{FR07,MRW11} to obtain stability and instability criteria of the zero solutions of
cross-diffusion systems.
From the numerical viewpoint different numerical schemes have been proposed to solve reaction-diffusion systems with nonlinear diffusion, see \cite{ABR11,BB04,GGJ03,GLS09,BB11,HHS94}.

The importance of the cross-diffusion, relatively to pattern formation, is extensively discussed in \cite{AP91,VE09}
from both the experimental and the theoretical point of view. In these papers the authors report many experiments of
interest to chemists where cross-diffusion effects can be significant: they obtain the minimal conditions for pattern
formation in the presence of linear cross-diffusion terms, demonstrating that relatively small values of
cross-diffusion parameters can lead to spatiotemporal pattern formation provided that the kinetics is sufficiently nonlinear.

The focus of this work is to describe the mechanisms of
pattern formation for the system (\ref{1.1}) with homogeneous
Neumann boundary conditions in a 2D domain.
The crucial difference with the 1D case analyzed in \cite{GLS12}
consists in the possibility that bifurcation occurs via a simple or
a multiple eigenvalue, see below and \cite{MM88,NM95,CMM97}.
We shall perform a weakly nonlinear analysis close
to the bifurcation state using the multiple scales analysis: this will give
the equations which rule the evolution of the pattern amplitude near the threshold.
A systematic approach to derive the normal forms
and the correspondent amplitude equations for flows at local bifurcations
can be found in \cite{CH93,Mann04,Hoy06,CroGreen,HHS94}. A comparison
between different methods of weakly nonlinear analysis for several
prototype reaction-diffusion equations is given in \cite{WMZ04}.

The paper is organized as follows: in Section \ref{Sec2}  a linear
stability analysis of the system \eqref{1.1} is performed and the
cross-diffusion is proved to be responsible for the initiation of
spatial patterns. In Section \ref{Sec3} the amplitude and the
form of the pattern close to the bifurcation threshold are investigated by using a weakly nonlinear multiple
scales analysis.
In particular, when the homogeneous steady state bifurcates to
spatial pattern at a simple eigenvalue, we derive the cubic and the quintic Stuart-Landau equation
which rules the evolution of the amplitude of the most unstable mode  in the
supercritical and subcritical case respectively.
In these cases the system supports patterns such rolls and squares.
On the other hand, when the bifurcation occurs via a double eigenvalue
more complex patterns arise due to the interaction of different modes
(for this reason they are called mixed mode patterns).
The corresponding evolution systems for the amplitudes of the pattern
are obtained and analyzed. A particular type of mixed mode patterns are
the hexagonal patterns, which arise when a resonance condition holds.
The evolution system for the amplitudes of the hexagonal patterns is proved to show bi-stability
and the phenomenon of hysteresis can be observed.

In all the considered cases the solutions predicted by the weakly nonlinear analysis are
compared with the numerical solutions of the original system.
Close to the threshold they show a good agreement.


\section{Cross-diffusion driven instability}\label{Sec2}
\setcounter{figure}{0}
\setcounter{equation}{0}

In this section we shall investigate the possibility of pattern appearance for the
system \eqref{1.1}.
In Subsection \ref{Sec2.1} we shall  determine the critical value for the bifurcation parameter
and the critical wavenumber via linear stability analysis. This will be done ignoring the geometry
of the domain and the role played by the boundary conditions.

This role will be considered in Subsection \ref{InstDeg} where we shall obtain the
range of the unstable wavenumbers of allowable patterns strictly depending on the domain geometry.
Since the degeneracy phenomenon can occur, the situation is much more involved than in the
1-D domain treated in \cite{GLS12} .

\subsection{Main results on the destabilization mechanism}\label{Sec2.1}

In order to stress the role played by the cross diffusion term in
the pattern forming process, the kinetics is chosen of the simplest
form, namely the competitive Lotka-Volterra, i.e. all $\gamma_{ij}>0$.
We shall only analyze the coexistence equilibrium:
\begin{equation}\label{equi_coe}
(u_0,v_0)\equiv
\left(\displaystyle\frac{\mu_1\gamma_{22}-\mu_2\gamma_{12}}{\gamma_{11}\gamma_{22}-\gamma_{12}\gamma_{21}},
\displaystyle\frac{\mu_2\gamma_{11}-\mu_1\gamma_{21}}{\gamma_{11}\gamma_{22}-\gamma_{12}\gamma_{21}}\right),
\end{equation}
as this is the only steady state relevant for pattern formation.
Therefore in this paper we shall assume the following conditions:
\begin{equation}\label{stab_cond_equi}
\mu_1\gamma_{22}-\mu_2\gamma_{12}>0, \qquad \mu_2\gamma_{11}-\mu_1\gamma_{21}>0, \qquad \gamma_{11}\gamma_{22}-\gamma_{12}\gamma_{21}>0 \, .
\end{equation}
The third condition (weak interspecific competition) is necessary for the stability of the equilibrium \eqref{equi_coe}.

Upon linearization of the system \eqref{1.1} in a neighborhood of $(u_0,v_0)$, namely:
\begin{equation}\label{2.1}
\dot{\textbf{w}}=\Gamma K\textbf{w}+D\nabla^2 \textbf{w},\ \ \
\qquad \textrm{where}\qquad
\textbf{w}=\left(\begin{array}{c}{u-u_0}\\
{v-v_0}\end{array}\right),
\end{equation}
and where:
\begin{eqnarray}\label{2.2}
K&=&\left(\begin{array}{cc}
-\gamma_{11}u_0  & -\gamma_{12}u_0\\
-\gamma_{21}v_0 & - \gamma_{22}v_0
\end{array}\right),\\\label{2.3}
\hskip1cm &&\nonumber\\
D&=&\left(\begin{array}{cc}
c_1+2a_1u_0+bv_0 & bu_0\\
b_2v_0 & c_2+2a_2v_0+b_2u_0
\end{array}\right),
\end{eqnarray}
we look for solutions in the form
$e^{i\mathbf{k}\cdot \mathbf{x}+ \sigma t}$.  Substitution in \eqref{2.1} leads to the following dispersion relation,
which gives the eigenvalue $\sigma$ as a function of the wavenumber $k=|\mathbf{k}|$:
\begin{equation}\label{2.4}
\sigma^2-g(k^2)\sigma+h(k^2)=0 \; ,
\end{equation}
where
\begin{equation}\label{h}
g(k^2)=k^2 \,{\rm tr}(D)-\Gamma \,{\rm tr}(K)\,,\qquad
h(k^2)={\rm det}(D)k^4+\Gamma q k^2+\Gamma^2 {\rm det}(K) \; ,
\end{equation}
and
\begin{equation}\label{2.5}
\begin{split}
q=&\ \gamma_{11}u_0(2a_2v_0+c_2)+\gamma_{22}v_0(2a_1u_0+c_1)
+bv_0(\gamma_{22}v_0-\gamma_{21}u_0)\\
&+b_2u_0(\gamma_{11}u_0-\gamma_{12}v_0)\; .
\end{split}
\end{equation}
For the Turing instability to be realized and spatial patterns to form,
$Re(\sigma)$ must be greater than zero for some $k\neq 0$. Since the polynomial
$g(k^2)>0, \forall k\neq 0$ (in fact ${\rm tr}(K)<0$ being $(u_0, v_0)$ stable
and ${\rm tr}(D)>0$), $Re(\sigma)$ will be positive for that $k\neq 0$ at which $h(k^2)<0$.
The steady state is marginally stable at some $k=k_c$ where:
\begin{equation}\label{ms}
{\rm min}(h(k_c^2))=0.
\end{equation}
As the minimum of $k$ is obtained when:
\begin{equation}\label{kc2}
k_c^2=-\frac{\Gamma q}{2\, {\rm det}(D^{b^c})},
\end{equation}
one has to require that $q$ can become negative.
In the above formula the matrix $D^{b^c}$ is the matrix $D$ defined in \eqref{2.3} evaluated at $b=b^c$.

From the expression \eqref{2.5} of $q$ where it is apparent that the first two terms are non negative, it
follows that the only potential destabilizing mechanism is the presence of the cross-diffusion terms.
Moreover only one of the last two terms in \eqref{2.5} can be negative, due to the conditions on positiveness
and stability of $(u_0, v_0)$. Therefore when $\gamma_{22}v_0-\gamma_{21}u_0<0$
(verified in the hyperbolic sector $P_1$  on the left of Fig. \ref{prima}), $b$ has a
destabilizing effect and $b_2$ acts as a stabilizer. Alternatively, when $\gamma_{11}u_0-\gamma_{12}v_0<0$
(region $P_2$ on the left of Fig. \ref{prima}), $b$ and $b_2$ exchange their role.
In the remainder of this paper we shall choose the kinetic parameter set in the first case and
the cross-diffusion coefficient $b$ as the bifurcation parameter.

Let us now write $q=-\alpha b+\beta$, where the positive quantities $\alpha$ and $\beta$ are defined as:
\begin{equation}
\begin{split}
\alpha&=v_0(\gamma_{21}u_0-\gamma_{22}v_0)\, ,\nonumber \\
\beta&=\gamma_{11}u_0(2a_2v_0+c_2)+\gamma_{22}v_0(2a_1u_0+c_1)
+b_2u_0(\gamma_{11}u_0-\gamma_{12}u_0)\, .
\end{split}
\end{equation}
Substituting $b=\beta/\alpha+\xi$ in the condition for the marginal stability \eqref{ms}
leads to the following equation for $\xi$:
\begin{equation}\label{2.7}
\begin{split}
\frac{\alpha^2}{4 {\rm det}(K)}\,\xi^2
-v_0(2a_2v_0+&c_2)\xi -[v_0\beta/\alpha(2a_2v_0+c_2)\\
+&(2a_1u_0+c_1)(2a_2v_0+b_2u_0+c_2)]=0.
\end{split}
\end{equation}
Then the critical bifurcation value is:
\begin{equation}\label{cbv}
b^c=\beta/\alpha+\xi^+,
\end{equation}
where $\xi^+$ is the positive root of equation \eqref{2.7} in such a way that $q<0$.

The results of this section can be summarized in the following theorem.

\begin{teo}
Suppose that $(u_0, v_0)$, as given in \eqref{equi_coe},  is a stable equilibrium for
the competitive kinetics of the system \eqref{1.1}, i.e. assume conditions \eqref{stab_cond_equi}
to be valid.

Moreover assume that $\gamma_{11}u_0-\gamma_{12}v_0<0$.
If:
$$
b>b^c,  
$$
where $b^c$ is defined in \eqref{cbv},
then $(u_0, v_0)$ is an unstable equilibrium for the reaction-diffusion
system \eqref{1.1}.
\end{teo}
An analogous statement would hold when $\gamma_{22}v_0-\gamma_{21}u_0<0$.

\begin{propos}
Suppose that $(u_0, v_0)$ is a stable equilibrium for
the competitive kinetics of the system \eqref{1.1}.

If $\gamma_{22}v_0-\gamma_{21}u_0<0$, the Turing parameter space of instability is the bounded region
$P_1$ on the left of Fig.\ref{prima}.
Alternatively, if $\gamma_{22}v_0-\gamma_{21}u_0<0$, the Turing parameter space of instability is the bounded region
$P_2$ .
\end{propos}

\subsection{Instability bands and degeneracy}\label{InstDeg}

\begin{figure}
\centering
\subfigure{\includegraphics[width=6.1cm, height=4.5cm]{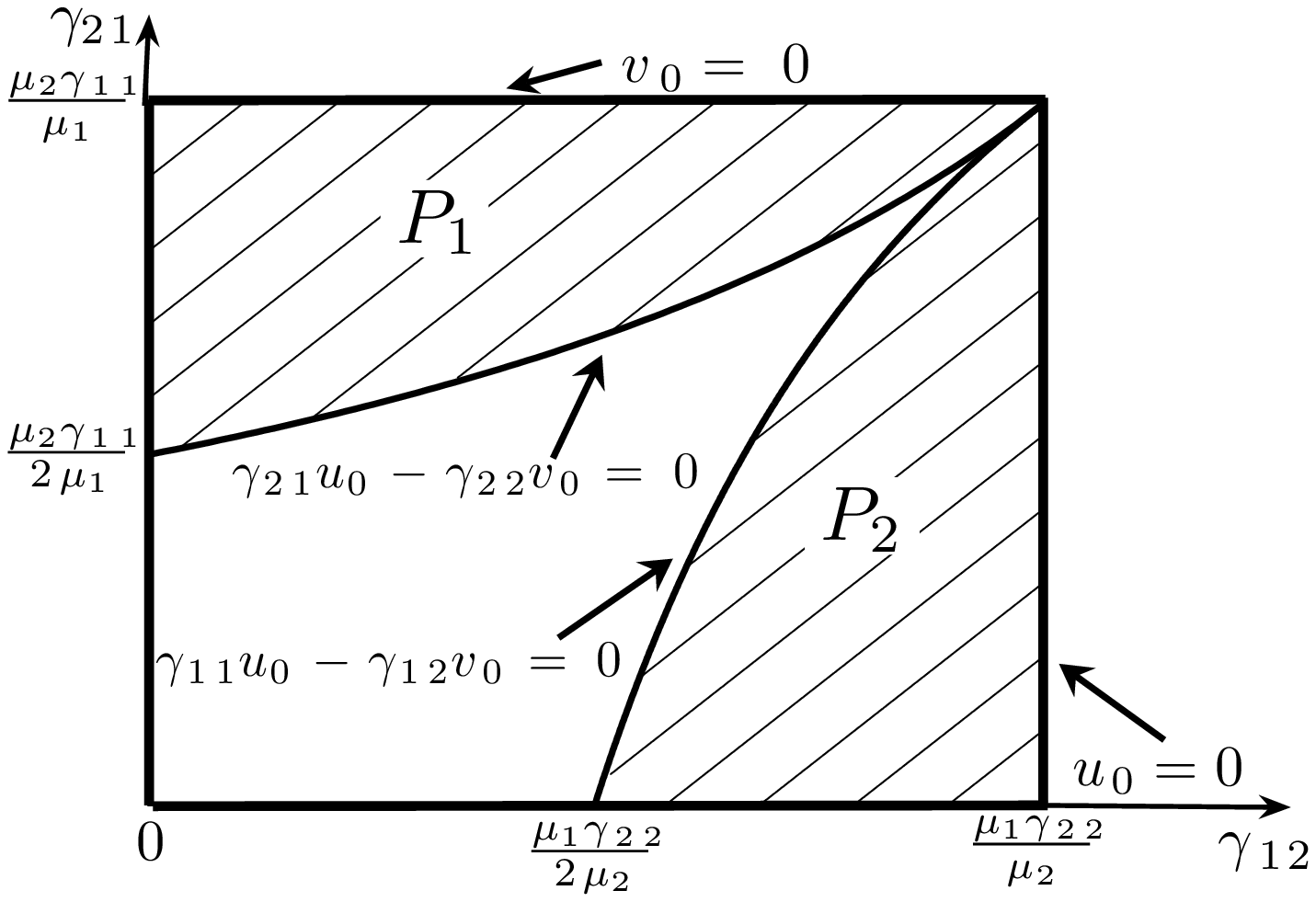}}
\subfigure{\includegraphics[width=6.cm, height=4.5cm]{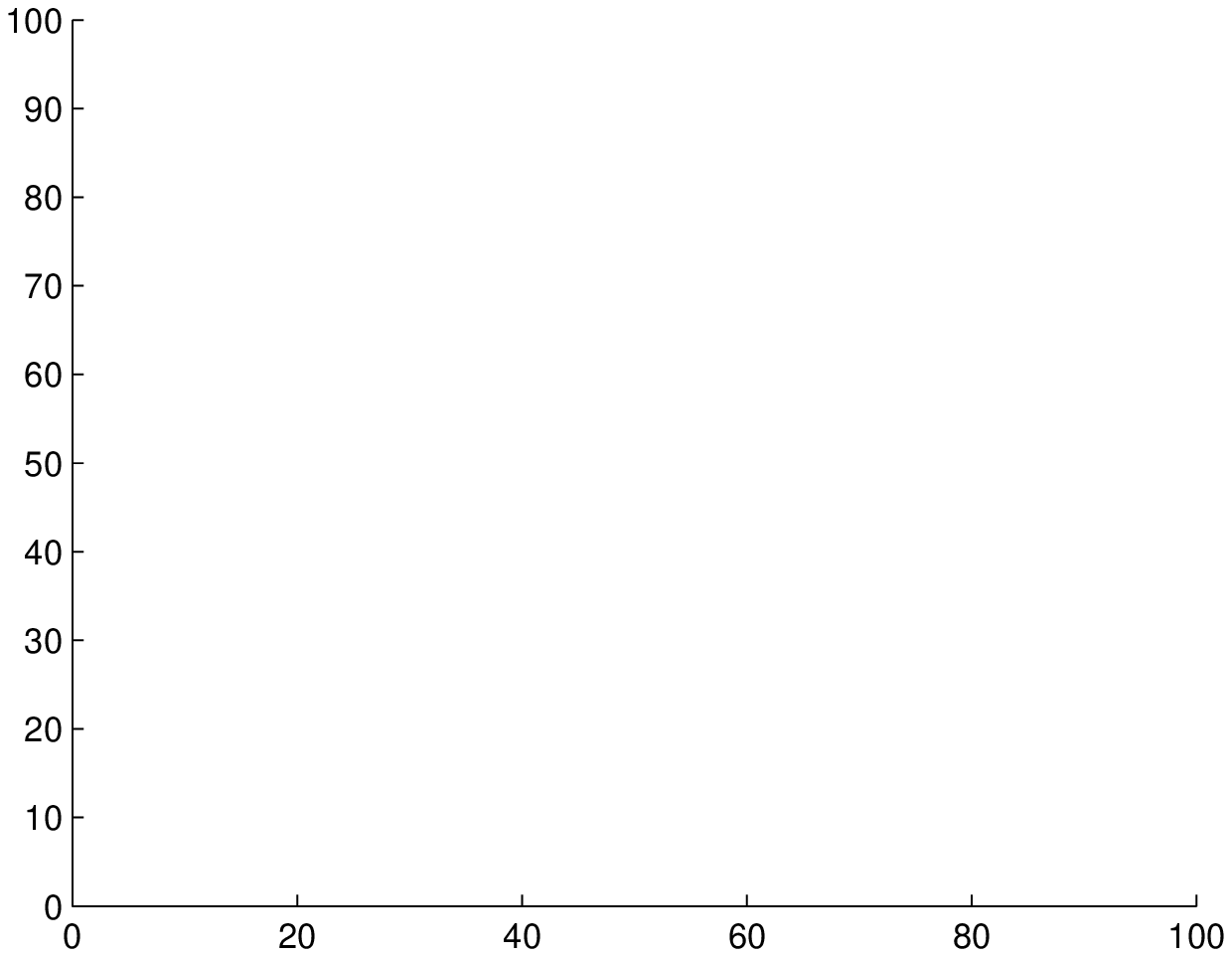}}
\caption{ {\em Left}: Pattern forming regions in the plane $(\gamma_{12}, \gamma_{21})$.
{\em Right}: Growth rate of the $k$-th mode.}\label{prima}
\end{figure}

When the domain is finite, the condition $b>b_c$ is not enough to see a pattern emerging.
In this case, in fact, $k_c$ might not be a mode admissible for the domain and the boundary condition.
However when $b>b^c$, there exists a range $(k_1^2, k_2^2)$ of unstable wavenumbers that make $h(k^2)<0$ and, correspondingly,
$Re(\sigma)>0$, see on the right of Fig.\ref{prima}.
It is easy to see that the extremes of the interval of unstable wavenumbers,
$k_1^2$ and $k_2^2$, where $h(k^2)=0$, are proportional to
$\Gamma$. It follows that $\Gamma$ must be big enough to find at least
one of the modes allowed by the Neumann boundary conditions within the interval $[k_1^2, k_2^2]$.

In a rectangular domain defined by $0<x<L_x$ and $0<y<L_y$, the solutions to the linear system \eqref{2.1}
with Neumann boundary conditions are:
\begin{eqnarray}\label{sol2d}
&&{\bf w}=\sum_{m,n\in \mathbb{N}}\mathbf{f}_{mn}\,e^{\sigma ({k}_{mn}^2)\,t}\,
\cos\left(\frac{m\pi}{L_x} \, x\right)\cos\left(\frac{m\pi}{L_y} \,y\right),\\ \label{k2d}
&&k_{mn}^2=\left(\frac{m\pi}{L_x}\right)^2+\left(\frac{n\pi}{L_y}\right)^2,
\end{eqnarray}
where $\mathbf{f}_{m,n}$ are  the Fourier coefficients of the initial conditions; the values  $\sigma(k_{mn}^2)$ are derived from the dispersion relation \eqref{2.4}.
The occurrence of a pattern emerging as $t$ increases, therefore depends on the existence of
mode pairs $(m,n)$ such that:
\begin{eqnarray}\label{con2d}
&&k_1^2<k^2\equiv \phi^2+\psi^2<k_2^2,\ {\rm where}\ \ \phi\equiv \frac{m\pi}{L_x},\ \
\psi\equiv  \frac{n\pi}{L_y},\\
&&\sigma(k^2)>0,
\end{eqnarray}
i.e. for $b>b^c$ and $\Gamma$ sufficiently large.
In what follows we shall restrict ourselves to the case when there is only
one unstable eigenvalue, admissible for the Neumann boundary conditions,
that falls within the band $(k_1, k_2)$ in the sense of Eq.\eqref{con2d}.
We shall denote this admissible eigenvalue with $\kcb$ to distinguish from the critical value $k_c$.

In a two dimensional domain, given $\kcb\in [k_1,k_2]$, one, two or more pairs $(m,n)$  may exist such that
the condition
\begin{equation}
\kcb^2=\phi^2+\psi^2=\left(\frac{m\pi}{L_x}\right)^2+\left(\frac{n\pi}{L_y}\right)^2 \label{kcbuguamn}
\end{equation}
is satisfied and in this case the eigenvalue $\sigma$ will have single, double or higher multiplicity respectively.
The multiplicity of the eigenvalue, and therefore the type of linear patterns we could expect,
strictly depends on the dimensions of the domain $L_x$ and $L_y$.

\begin{figure}
\begin{center}
\epsfxsize=5.65cm\epsfbox{vuoto.eps} \epsfxsize=5.65cm \epsfbox{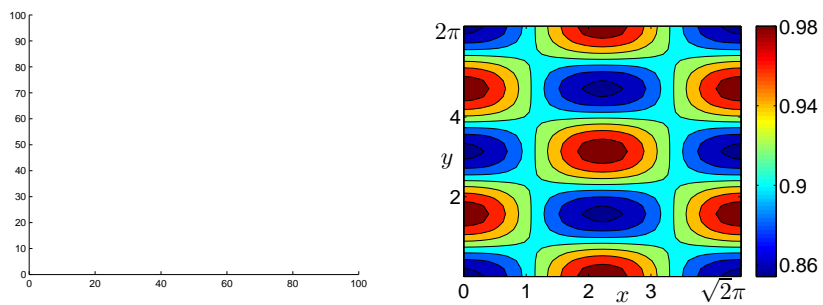}
\end{center}
\caption{{\em Left}: The species $u$. {\em Right}: The species $v$. The parameters are $\mu_1=1.2$, $\mu_2=1$, $\gamma_{11}=0.5$, $\gamma_{12}=0.4$, $\gamma_{21}=0.38$, $\gamma_{22}=0.4$, $a_1=0.01$,
$a_2=0.001$, $c_1=0.1$, $c_2=0.2$, $\Gamma=28.05$, $b_2=1.1$, $b=7.264>b^c=7.192$.
}
\label{seconda}
\end{figure}

In Fig.\ref{seconda} we show the pattern which forms starting from
an initial datum which is a random periodic perturbation about the steady state
$(u_0,v_0)$. For the  parameters we have picked for this simulation, one has $(u_0,v_0)\approx (1.67, 0.92)$
while the critical value of the bifurcation parameter is $b^c=7.192$.
In the rectangular domain with $L_x=\sqrt{2}\pi$ and $L_y=2\pi$ only the mode $\kcb^2=6$
is admitted by the boundary conditions. The eigenvalue predicted by the linear analysis
is single: in fact there only exists the  pair $(2,4)$ which satisfies the condition \eqref{k2d}.

All the numerical simulations showed in this paper are performed by using spectral methods.
Here we have employed $32$ modes both in the $x$ and in the $y$ axis. However the use
of a higher number of modes in the scheme (we tested the method up to $128$ modes) does not
appreciably affect the results.
Notice that in all the figures representing the spectrum of the solutions, for a better presentation of the results, the amplitude of the zero mode, corresponding to the equilibrium solution, has been set equal to zero.

\begin{figure}
\begin{center}
\epsfxsize=5.65cm \epsfbox{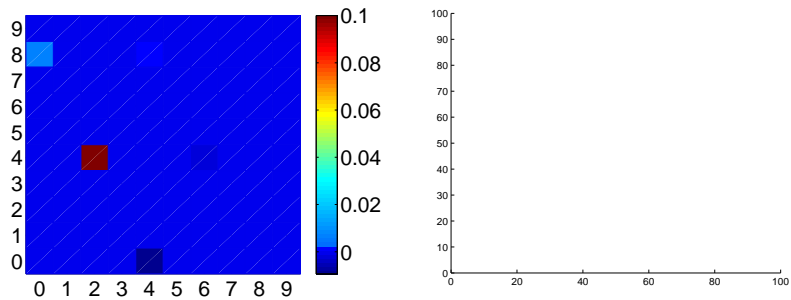}\epsfxsize=5.65cm \epsfbox{vuoto.eps}
\end{center}
\caption{Spectrum of the numerical solutions in Fig. \ref{seconda}.}
\label{seconda_bis}
\end{figure}

\section{Weakly nonlinear analysis}\label{Sec3}
\setcounter{equation}{0}
\setcounter{figure}{0}

In this section a weakly nonlinear analysis is carried out to obtain the amplitude equations describing the dynamics near the critical bifurcation state 
.
The method of multiple scales (as introduced in \cite{NW69}), together
with an asymptotic analysis of the system close to its marginal stability are employed
to determine the near-critical bifurcation structure of the patterns \cite{CH93}
.

Near the threshold the amplitude of the
pattern evolves on a slow temporal scale, and therefore we shall introduce new scaled coordinates
separating the fast time $t$ and slow time  e.g. $T=\varepsilon t$; here
the control parameter $\varepsilon$ will
measure the distance of the system from bifurcation, see \eqref{4.5} below.
The solution of the original system \eqref{1.1} is written as an
expansion in $\varepsilon$ and the leading order term
of this expansion is shown to be the product of
the basic pattern (the critical solution of the linearized system
\eqref{2.1}) and a slowly varying amplitude (see \cite{HHS94,Hoy06}).
We shall confine ourselves to patterns which are modulated in time but not in space,
so that in our analysis we do not take into account the slow spatial scale.

If one defines the linear operator:
\begin{equation}\label{3.2}
\mathcal{L}^b=\Gamma \, K + D^{b} \nabla^2,
\end{equation}
where $K$ and $D^{b}$ are given in \eqref{2.2} and \eqref{2.3}, and if one introduces the following
bilinear operators acting on $(\mathbf{x}, \mathbf{y})$ with $\textbf{x}\equiv(x^u,x^v)$ and
$\textbf{y}\equiv(y^u,y^v)$:
\begin{eqnarray}\label{opK}
\mathcal{Q}_K(\textbf{x},\textbf{y})&=&\Gamma \left(
\begin{array}{c}
-2\gamma_{11}x^uy^u- \gamma_{12}(x^u y^v+x^v y^u) \\
-2\gamma_{22}x^vy^v- \gamma_{21}(x^uy^v+x^vy^u)
\end{array}\right),\\
\ \nonumber\\\label{opD}
\mathcal{Q}_D^{b}(\textbf{x},\textbf{y})&=&\left(
\begin{array}{c}
2a_1x^uy^u+b(x^uy^v+x^vy^u) \\
2a_2x^vy^v+b_2(x^uy^v+x^vy^u)
\end{array}\right),
\end{eqnarray}
the original system \eqref{1.1} can be rewritten separating the linear
and the nonlinear part as follows:
\begin{equation}\label{3.1}
\partial_t \textbf{w}= \mathcal{L}^{b} \textbf{w}+
\frac{1}{2}\mathcal{Q}_K(\textbf{w},\textbf{w})
+\frac{1}{2}\nabla^2 \mathcal{Q}_D^{b}(\textbf{w},\textbf{w}),
\end{equation}
with $\textbf{w}$ defined as in \eqref{2.1}.

Let us introduce the multiple time scales:
\begin{equation}\label{4.3}
t=\frac{T_1}{\varepsilon}+\frac{T_2}{\varepsilon^2}+\frac{T_3}{\varepsilon^3}+\frac{T_4}{\varepsilon^4}+\dots
\end{equation}
and expand accordingly both the solution ${\bf w}$ and the bifurcation
parameter $b$:
\begin{eqnarray}\label{4.4}
{\bf w}&=&\varepsilon {\bf w}_1+\varepsilon^2 {\bf w}_2+\varepsilon^3 {\bf w}_3+\varepsilon^4 {\bf w}_4+O(\varepsilon^5),\\
b&=&b^c+\varepsilon b^{(1)}+\varepsilon^2 b^{(2)}+\varepsilon^3 b^{(3)}+\varepsilon^4 b^{(4)}+O(\varepsilon^5).\label{4.5}
\end{eqnarray}
Expansion \eqref{4.5} can be considered as the definition of the smallness parameter.
In the rest of this paper, we shall always measure the distance from the threshold using, as unit, the
critical value $b^c$.
This means that, when different from zero, we shall set $b^{(i)}=b^{c}$.
Substituting \eqref{4.3}-\eqref{4.5} into the full system (\ref{1.1}),
the following sequence of linear equations for $\mathbf{w}_i$ is obtained:
\begin{eqnarray}\label{4.6bis}
O(\varepsilon)&:&\\\nonumber
&&\mathcal{L}^{b^c} {\bf w}_1=\mathbf{0},\\ \nonumber
\
\\ \label{4.7}
O(\varepsilon^2)&:&\\\nonumber
&&\mathcal{L}^{b^c} {\bf
w}_2=\mathbf{F}=\frac{\partial {\bf w}_1}{\partial T_1}
-\frac{1}{2}\left(\mathcal{Q}_K+\nabla^2\mathcal{Q}_D^{b^c}\right)(\mathbf{w}_1,
\mathbf{w}_1)\\\nonumber
&&\quad\quad\quad-b^{(1)}\left(\begin{array}{cc}v_0 & u_0\\0 & 0\end{array}\right)\nabla^2{\bf w}_1,\\\nonumber
\ \\\label{4.10}
O(\varepsilon^3)&:&\\\nonumber
&&\mathcal{L}^{b^c} {\bf w}_3=\mathbf{G}=\frac{\partial {\bf
w}_1}{\partial T_2}+\frac{\partial {\bf w}_2}{\partial
T_1}-\left(\mathcal{Q}_K+\nabla^2\mathcal{Q_D}^{b^c}\right)(\mathbf{w}_1,
\mathbf{w}_2)\\\nonumber
&&\quad\quad\quad-b^{(1)}\nabla^2\left(\begin{array}{c} u_1v_1\\
0\end{array}\right)-\left(\begin{array}{cc}v_0 & u_0\\0 & 0\end{array}\right)\left(b^{(1)}\nabla^2{\bf w}_2+b^{(2)}\nabla^2{\bf w}_1\right),\\\nonumber
\ \\\label{4.10ep4}
O(\varepsilon^4)&:&\\\nonumber
&&\mathcal{L}^{b^c} {\bf w}_4=\mathbf{H}=\frac{\partial {\bf
w}_1}{\partial T_3}+\frac{\partial {\bf w}_2}{\partial
T_2}+\frac{\partial {\bf w}_3}{\partial
T_1}-\left(\mathcal{Q}_K+\nabla^2\mathcal{Q_D}^{b^c}\right)(\mathbf{w}_1,
\mathbf{w}_3)\\\nonumber
&&\quad\quad\quad-\frac{1}{2}\left(\mathcal{Q}_K+\nabla^2\mathcal{Q_D}^{b^c}\right)(\mathbf{w}_2,
\mathbf{w}_2)-b^{(1)}\nabla^2\left(\begin{array}{c} u_1v_2+u_2v_1\\
0\end{array}\right)\\\nonumber
&&\quad\quad\quad-\left(\begin{array}{cc}v_0 & u_0\\0 & 0\end{array}\right)\left(b^{(1)}\nabla^2{\bf w}_3+b^{(2)}\nabla^2{\bf w}_2+b^{(3)}\nabla^2{\bf w}_1\right)\\\nonumber
&&\quad\quad\quad-b^{(2)}\nabla^2\left(\begin{array}{c} u_1v_1\\
0\end{array}\right),\\\nonumber
\ \\\label{4.10ep5}
O(\varepsilon^5)&:&\\\nonumber
&&\mathcal{L}^{b^c} {\bf w}_4=\mathbf{P}=\frac{\partial {\bf
w}_1}{\partial T_4}+\frac{\partial {\bf w}_2}{\partial
T_3}+\frac{\partial {\bf w}_3}{\partial
T_2}+\frac{\partial {\bf w}_4}{\partial
T_1}\\\nonumber
&&\quad\quad\quad-\left(\mathcal{Q}_K+\nabla^2\mathcal{Q_D}^{b^c}\right)(\mathbf{w}_1,
\mathbf{w}_4)-\left(\mathcal{Q}_K+\nabla^2\mathcal{Q_D}^{b^c}\right)(\mathbf{w}_2,
\mathbf{w}_3)
\\\nonumber
&&\quad\quad\quad-\left(\begin{array}{cc}v_0 & u_0\\0 & 0\end{array}\right)\left(b^{(1)}\nabla^2{\bf w}_4
+b^{(2)}\nabla^2{\bf w}_3+b^{(3)}\nabla^2{\bf w}_2
+b^{(4)}\nabla^2{\bf w}_1\right)\\\nonumber
&&\quad\quad\quad-b^{(1)}\nabla^2\left(\begin{array}{c} u_1v_3+u_2v_2+u_3v_1\\
0\end{array}\right)-b^{(2)}\nabla^2\left(\begin{array}{c} u_1v_2+u_2v_1\\
0\end{array}\right)\\\nonumber
&&\quad\quad\quad-b^{(3)}\nabla^2\left(\begin{array}{c} u_1v_1\\
0\end{array}\right).
\end{eqnarray}

The solution of the linear problem \eqref{4.6bis} satisfying the Neumann
boundary conditions is given by:
\begin{equation}\label{4.6}
{\bf w}_1=\sum_{i=1}^m A_i(T_1, T_2)\bfrho\cos(\phi_i x)\cos(\psi_i
y)\, ,
\end{equation}
where $m$ is the multiplicity of the eigenvalue, 
$A_i$ are the slowly varying amplitudes (still arbitrary at this level),
while $\bfrho$, which is defined up to a constant, is explicitly given by the following formulas:
\begin{equation}\label{emme}
\bfrho \in \mbox{Ker}(\Gamma K-k_c^2D^{b^c}), \qquad \bfrho =  \left(\begin{array}{c} 1 \\M \end{array}\right) \, ,
\qquad \mbox{with} \quad
 M\equiv\frac{-D^{b^c}_{21}{k}_c^2+\Gamma K_{21}}{D^{b^c}_{22}{k}_c^2-\Gamma K_{22}},
\end{equation}
where $D^{b^c}_{ij}, K_{ij}$ are the $i,j$-entries of
the matrices $D^{b^c}$ and $K$.

We shall restrict our analysis to cases where the multiplicity is $m=1$ or $2$.

\subsection{Simple eigenvalue, $m=1$}

If the eigenvalue is simple, i.e. $m=1$, the solution \eqref{4.6} at the lowest order reduces to:
\begin{equation}
{\bf w}_1=A(T_1, T_2)\bfrho\cos(\phi_1 x)\cos(\psi_1
y)\, .
\end{equation}
Substituting this expression into the linear equation \eqref{4.7},
the vector $\mathbf{F}$ is made orthogonal to the kernel of the adjoint of $\mathcal{L}^{b^c}$ simply
by imposing $T_1=0$ and $b^{(1)}=0$.
The solution of (\ref{4.7}) can therefore be obtained (see \eqref{w2_m1}) and substituted into the
linear problem \eqref{4.10} at order $\varepsilon^3$.
The vector $\mathbf{G}$, given by \eqref{G_m1}, contains secular terms,
and therefore it does not automatically satisfy the Fredholm condition.
By imposing the compatibility condition  the following Stuart-Landau equation for the amplitude $A(T_2)$ is finally obtained:
\begin{equation}\label{4.11}
\frac{dA}{dT_2}=\sigma A-L A^3\, ,
\end{equation}
where the expression of $\sigma$ and $L$ are in terms of the parameters of the system \eqref{1.1} 

\begin{figure}
\begin{center}
\epsfxsize=5.65cm\epsfbox{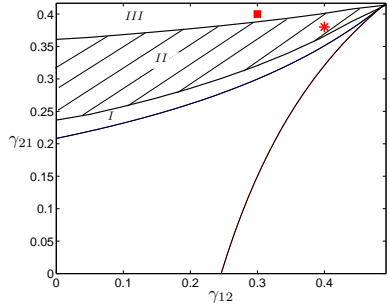}
\end{center}
\caption{Within the pattern forming region, the zones of subcritical (I and III) and supercritical (II, dashed) bifurcation are drawn. The parameters
are $\mu_1=1.2$, $\mu_2=1$, $\gamma_{11}=0.5$, $\gamma_{22}=0.41$,
$c_1=c_2=0.2$, $a_1=a_2=0.1$, $b_2=0.154$.}\label{regionsubsuper}
\end{figure}

In the pattern-forming region the growth rate coefficient $\sigma$ is always positive.
Therefore one can distinguish two cases for the qualitative dynamics of the Stuart-Landau equation \eqref{4.11}.
The supercritical case, when $L>0$, and the subcritical case, when $L<0$.
Since the expression for $L$ as a function of all the parameters of the original system is quite involved,
we numerically determine the curves in the space $(\gamma_{12}, \gamma_{21})$ across which $L$ changes
its sign (all the other parameters being fixed).
These curves divide the pattern forming domain into three regions, as shown in Fig.\ref{regionsubsuper}: region II (dashed) corresponds to the supercritical bifurcation, while regions I and III correspond to
the subcritical case.

\subsubsection{The supercritical case}
When $\sigma$ and $L$ are both positive, the solution of the Stuart-Landau equation evolves towards the stable stationary state $A_{\infty}=\sqrt{\sigma/L}$. Therefore:

\begin{teo}\label{teosup}
Assume that:
\begin{enumerate}
\item
$\ep^2=(b-b^c)/b^c$ is small enough so that
the uniform steady state $(u_0,v_0)$ in \eqref{equi_coe} is unstable
to modes corresponding only to the eigenvalue $\kcb$;
\item
there exists only one couple of integers $(m,n)$ such that:
\[\kcb^2\equiv \phi^2+\psi^2\ {\rm where}\ \ \phi\equiv \frac{m\pi}{L_x},\ \
\psi\equiv  \frac{n\pi}{L_y},
\]
\item
the Landau coefficient $L$  is greater than zero.
\end{enumerate}

Then the emerging solution of the reaction-diffusion
system \eqref{1.1} is given by:
\begin{equation}\label{sol_m1}
\mathbf{w}=\varepsilon \bfrho A_{\infty} \cos(\phi x)\cos(\psi y)+O(\varepsilon^2),
\end{equation}
where $A_{\infty}$ is the stable stationary state of the Stuart-Landau equation \eqref{4.11},
and $\bfrho$ is given in \eqref{emme}.
\end{teo}

In the first numerical test we choose the set of parameters in such a way that the
only unstable mode allowed by the boundary conditions is $\kcb^2=8$ and a supercritical
bifurcation arises (i.e. $\sigma$ and $L$ are both positive). In particular the parameters are chosen as in Fig.\ref{regionsubsuper} at the point marked with an asterisk.

\begin{figure}
\begin{center}
\epsfxsize=5.65cm\epsfbox{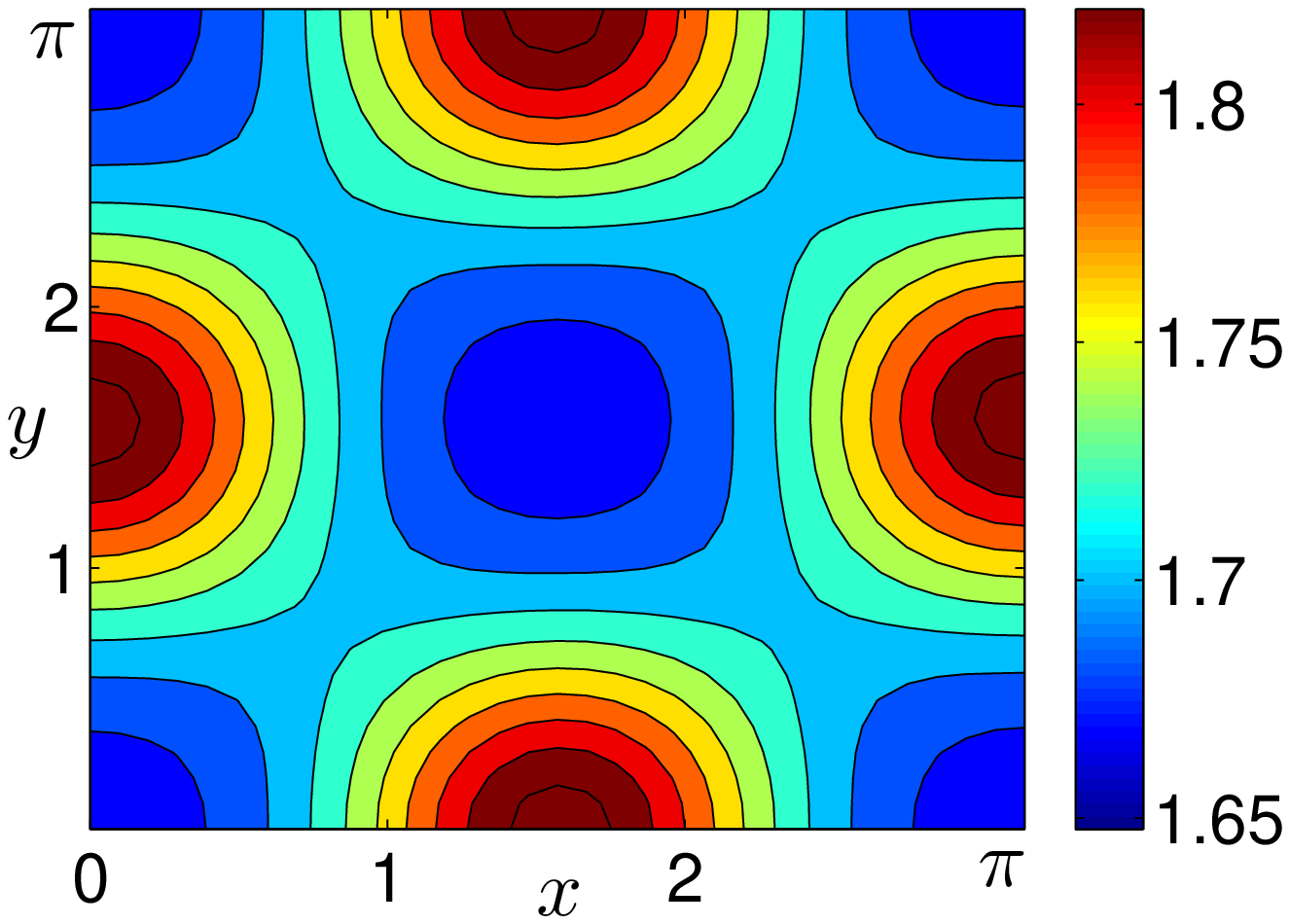} \epsfxsize=5.65cm
\epsfbox{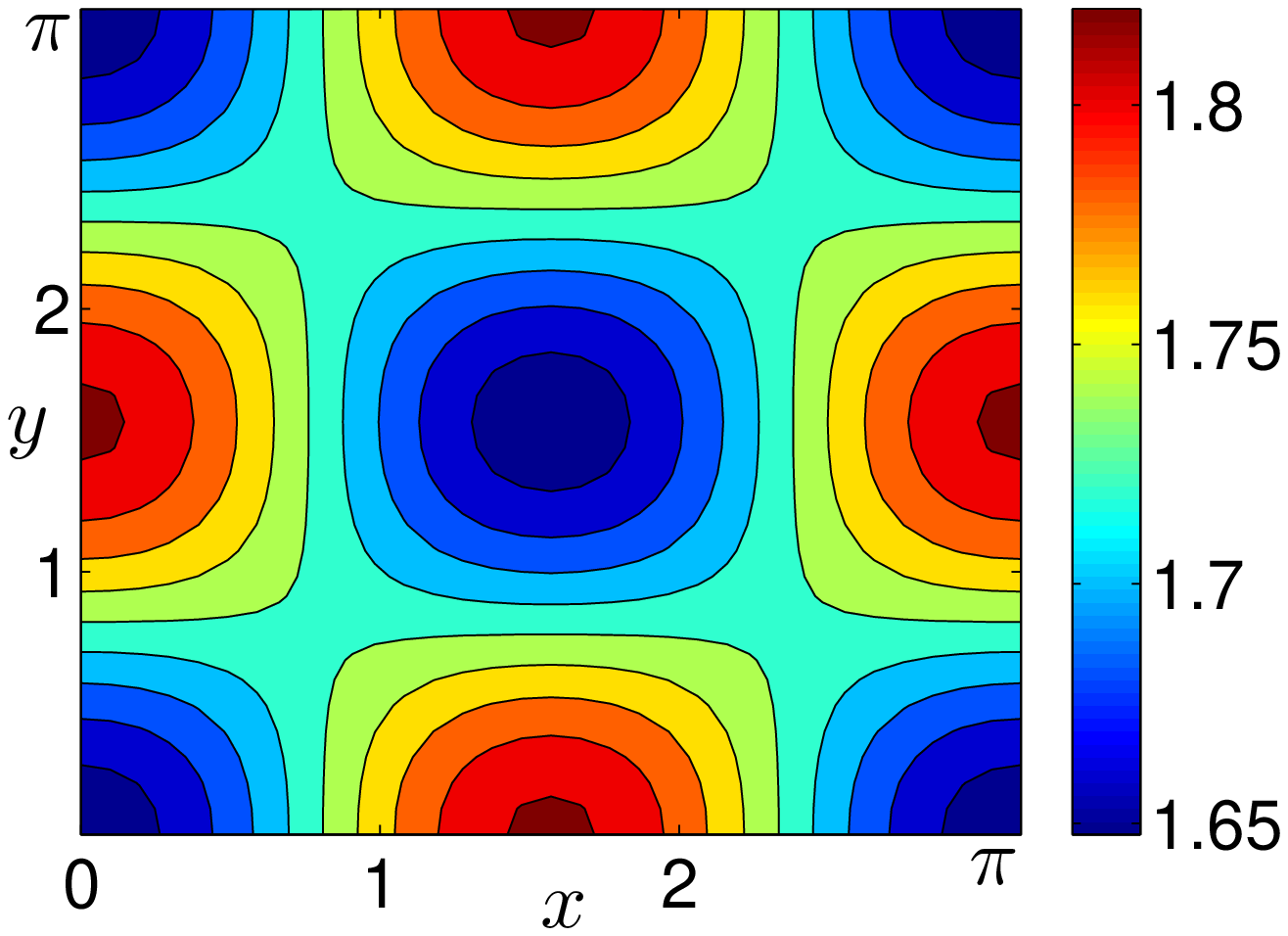}
\end{center}
\caption{\textit{Supercritical case}. Comparison between the numerical solution (on the left)
and the weakly nonlinear first order approximation of the solution
(on the right). The system parameters are chosen as in Fig.\ref{regionsubsuper},
at the point $\gamma_{12}=0.4$, $\gamma_{21}=0.38$, marked with an asterisk.
Moreover, $\Gamma=40.75$,  $b^c=5.37$, $\varepsilon=0.1$,
$b=(1+0.1^2)\,b^c=5.43$.}\label{nona}
\end{figure}

On a square
domain with dimensions $L_x=L_y=\pi$, there exists only the pair
$(m,n)=(2,2)$ such that formula
\eqref{k2d} is satisfied. In this case the system supports
square patterns.
The first order approximation of the
solution predicted by the weakly nonlinear analysis reads:
\begin{equation}
\textbf{w}=\varepsilon A_{\infty}\bfrho \cos(2x)cos(2y)+O(\varepsilon^2),
\end{equation}
which shows a good agreement with the numerical solution of
the full system (\ref{1.1}),  see Fig.\eqref{nona}.
We have verified that the error in predicting the amplitude is $O(\varepsilon^2)$.

\subsubsection{The subcritical case}
For certain values of the parameters (see regions I and III in Fig.\ref{regionsubsuper}),
the Landau coefficient $L$ has a
negative value. In these cases Eq. \eqref{4.11} is not able to capture
the amplitude of the pattern. This is a typical situation where
the transition occurs {\em via} a subcritical bifurcation. To predict the amplitude of the pattern, one
needs to push the weakly nonlinear expansion to a higher order
(for a general discussion on the relevance of the higher order
amplitude expansions in the study of subcritical bifurcations, see
the recent \cite{BMS09} and references therein).

Performing the weakly nonlinear analysis up to $O(\varepsilon^5)$
one obtains the following quintic Stuart-Landau equation
for the amplitude $A$:
\begin{equation}\label{quintic_SL}
\frac{d A}{d
T_2}=\bar{\sigma}A-\bar{L}A^3+\bar{Q}A^5\, .
\end{equation}
%
One can summarize the analysis as:

\textbf{WNL analysis results in the non degenerate subcritical case}
\emph{
Assume that the hypotheses (1) and (2) of Theorem \ref{teosup} hold and that
\begin{enumerate}
\item[(3)] the Landau coefficient $L$  is negative;
\item[(4)] the coefficient  $\bar{Q}$ is positive.
\end{enumerate}
Then the emerging solution of the reaction-diffusion
system \eqref{1.1} is given by:
\begin{equation}\label{sol_m1quint}
\mathbf{w}=\varepsilon \bfrho A_{\infty} \cos(\phi x)\cos(\psi y)+O(\varepsilon),
\end{equation}
where $A_{\infty}$ is a stable stationary state of the quintic Stuart-Landau equation \eqref{quintic_SL}.
}

It is important to notice that, given that the coefficient $\bar{Q}=O(\ep^2)$ while $\bar{\sigma}$ and $\bar{L}$ are
$O(1)$, the equilibria $A_\infty=O(\ep^{-1})$. This means that the emerging
pattern is an $O(1)$ perturbation of the equilibrium, which contradicts the basic assumption of the
perturbation scheme \eqref{4.4}.
In the subcritical case one should therefore expect significant quantitative discrepancies between the full system
results and the predictions of the WNL analysis.
Nevertheless in our simulation we have always found a fairly good qualitative agreement of the approximation \eqref{sol_m1quint}
with the solution of the full system; most importantly we have seen that the bifurcation diagram constructed using
\eqref{quintic_SL} is able to predict very well phenomena like bistability and hysteresis cycle shown also by the full system.

\begin{figure}
\begin{center}
\epsfxsize=5.8cm\epsfbox{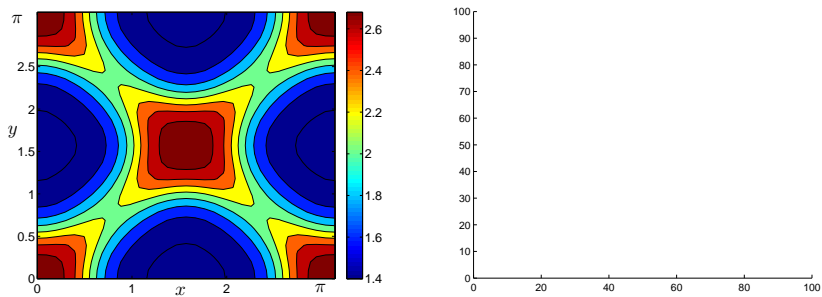}\epsfxsize=5.8cm\epsfbox{vuoto.eps}
\end{center}
\caption{\textit{Subcritical case}. Comparison between the numerical solution of
(\ref{1.1}) with $\epsilon\approx 0.32$ (Left) and the third order weakly
nonlinear approximated solution (Right) based on the quintic Stuart Landau equation \eqref{quintic_SL}.
The parameters are chosen as in the Fig.~\ref{regionsubsuper} at the point $\gamma_{12}=0.3$,
$\gamma_{21}=0.4$, denoted with a square. Moreover, $\Gamma\approx 32$, $b^c=5.63$, $\kcb^2=8$, $\ep=0.1$, $b\approx 6.25$. }\label{compsub1amp}
\end{figure}

In Fig.~\ref{compsub1amp} we show a comparison between the
numerical solution of the system (\ref{1.1}) and the weakly nonlinear
approximation for the choice of the parameters corresponding to the point denoted with a square in Fig.\ref{regionsubsuper}. In Table 3.1 the values of the amplitudes of the most excited modes of the numerical
and the approximated solutions furnished by the weakly nonlinear analysis are compared.

\begin{figure}
\begin{center}
\epsfxsize=7.65cm\epsfbox{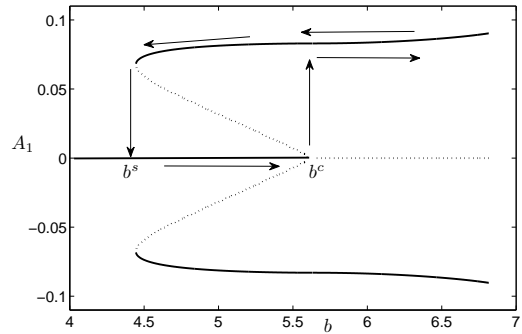}
\end{center}
\caption{The bifurcation diagram of the quintic Stuart-Landau equation \eqref{quintic_SL}.
The parameters are chosen as in Fig.~\ref{compsub1amp}. }\label{bif1d}
\end{figure}

\begin{figure}
\begin{center}
\epsfxsize=5.35cm \epsfbox{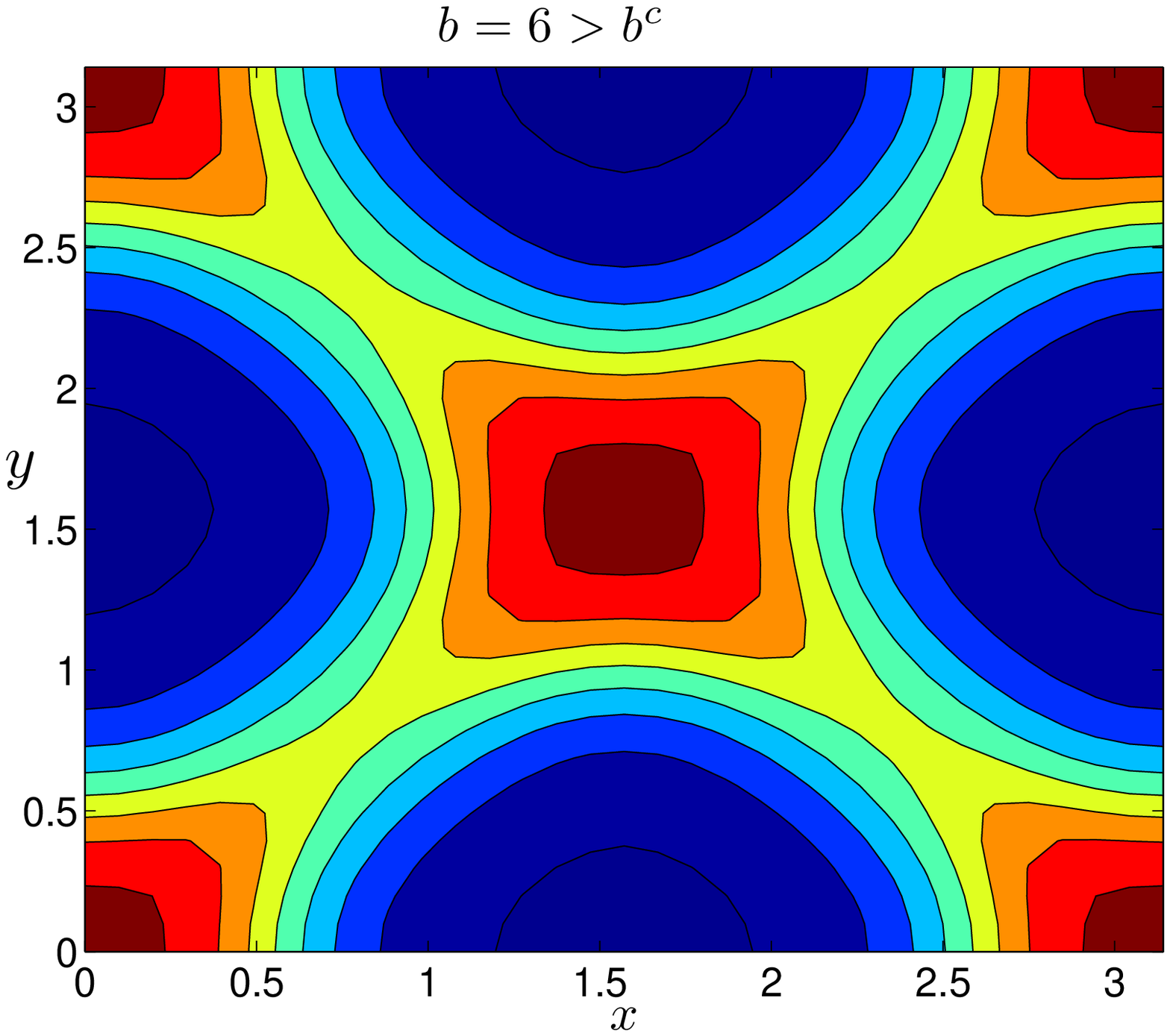} \epsfxsize=5.15cm\epsfbox{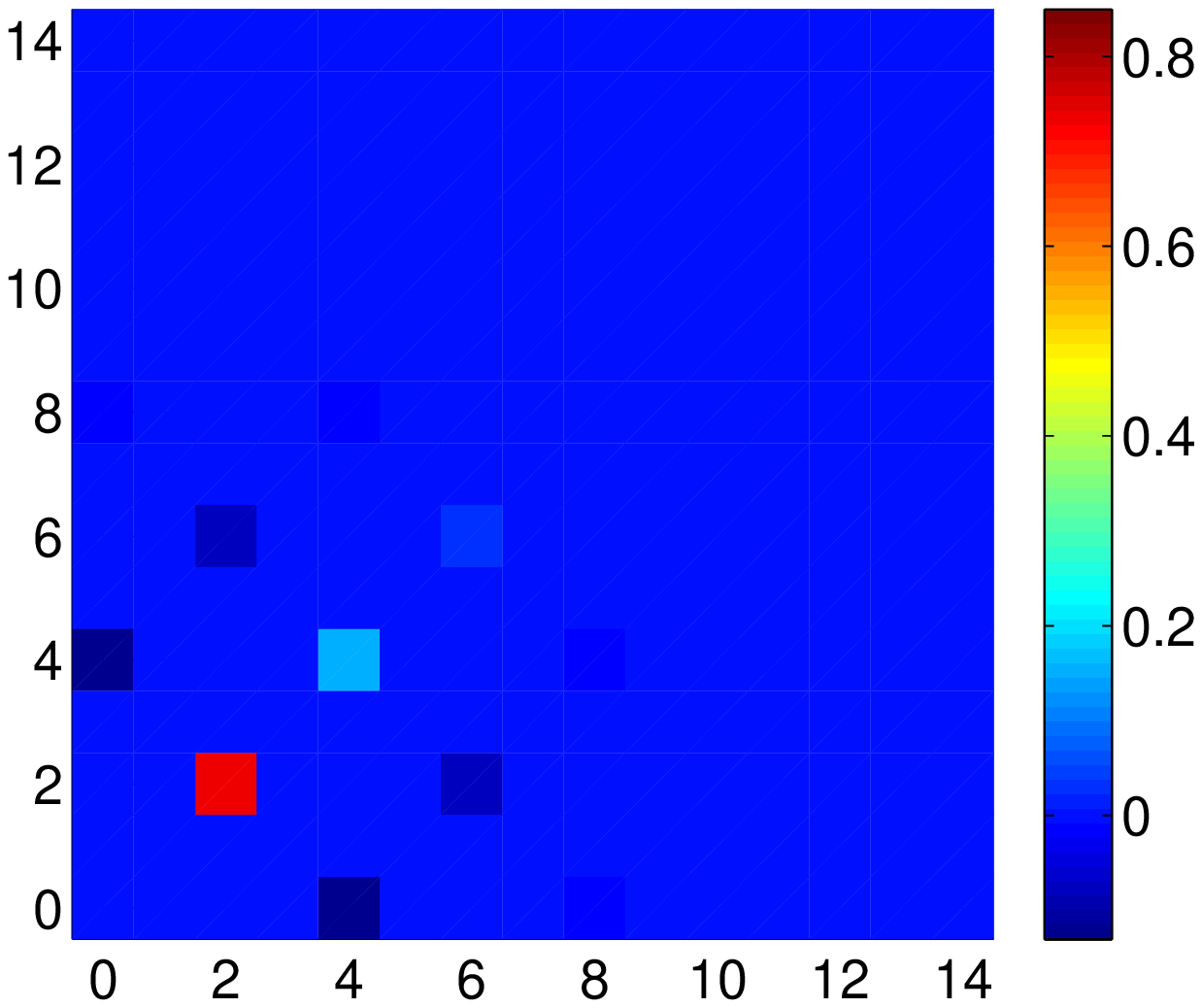}
\epsfxsize=5.35cm \epsfbox{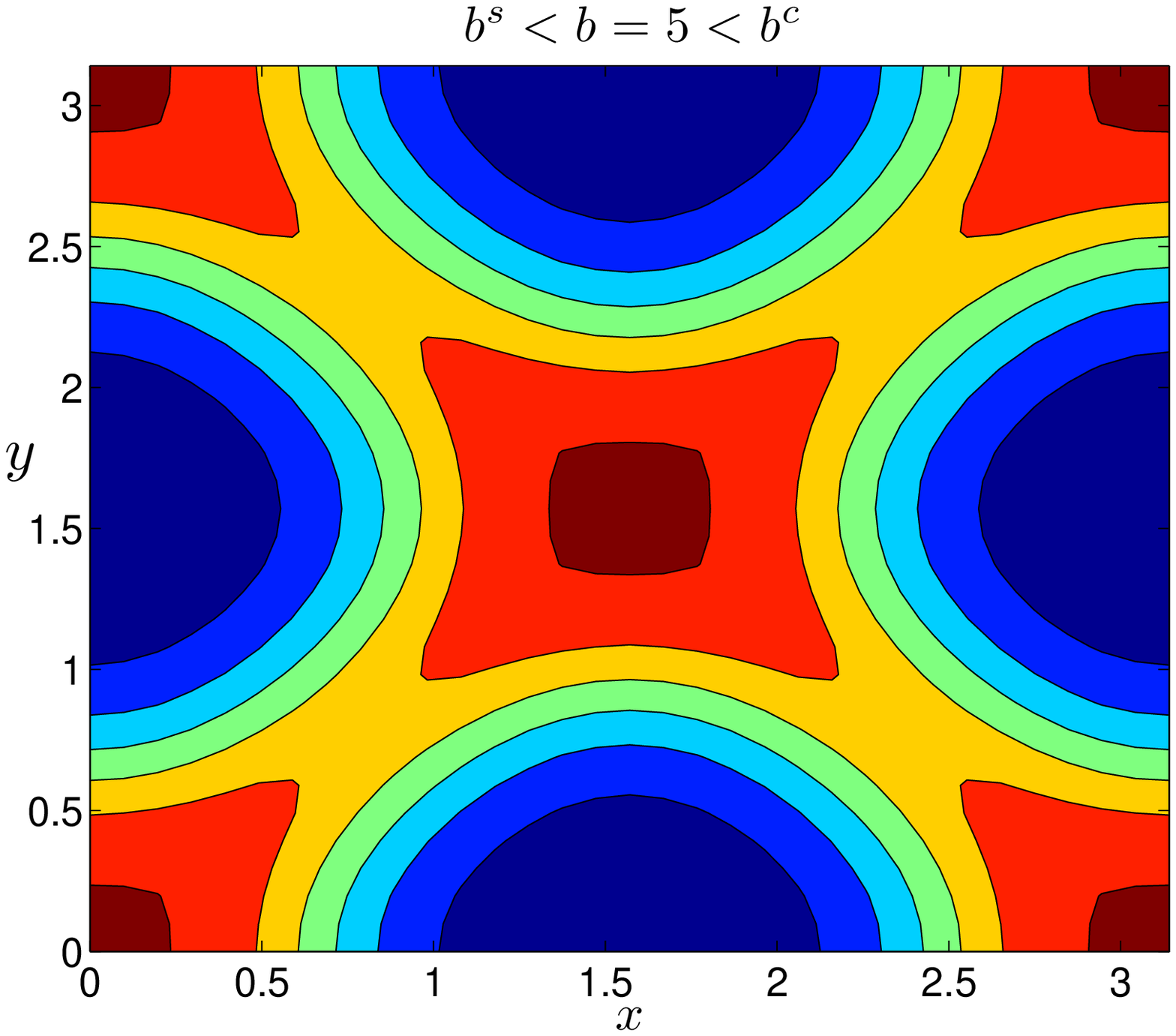} \epsfxsize=5.15cm\epsfbox{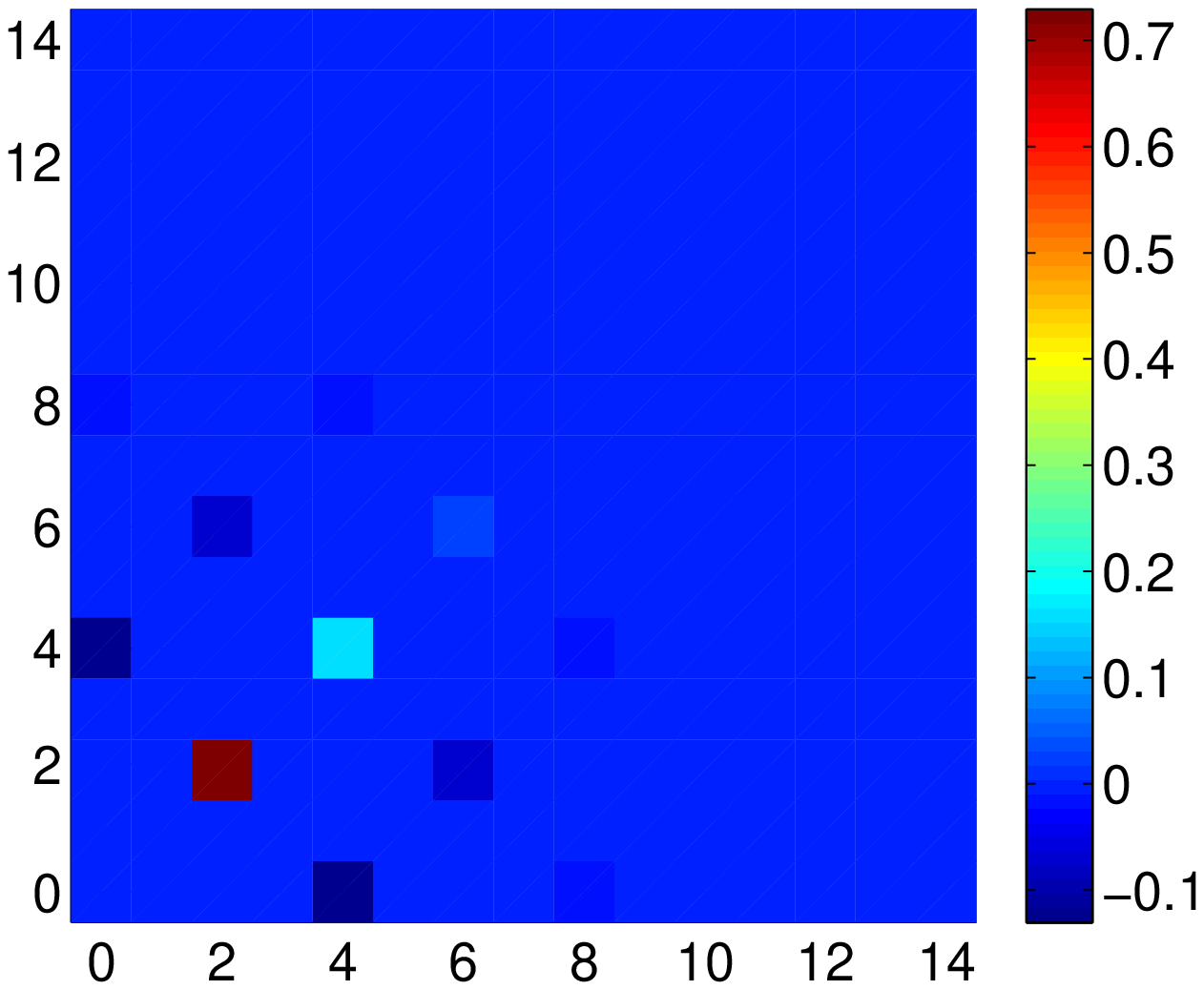}
\epsfxsize=5.35cm \epsfbox{vuoto.eps} \epsfxsize=5.15cm\epsfbox{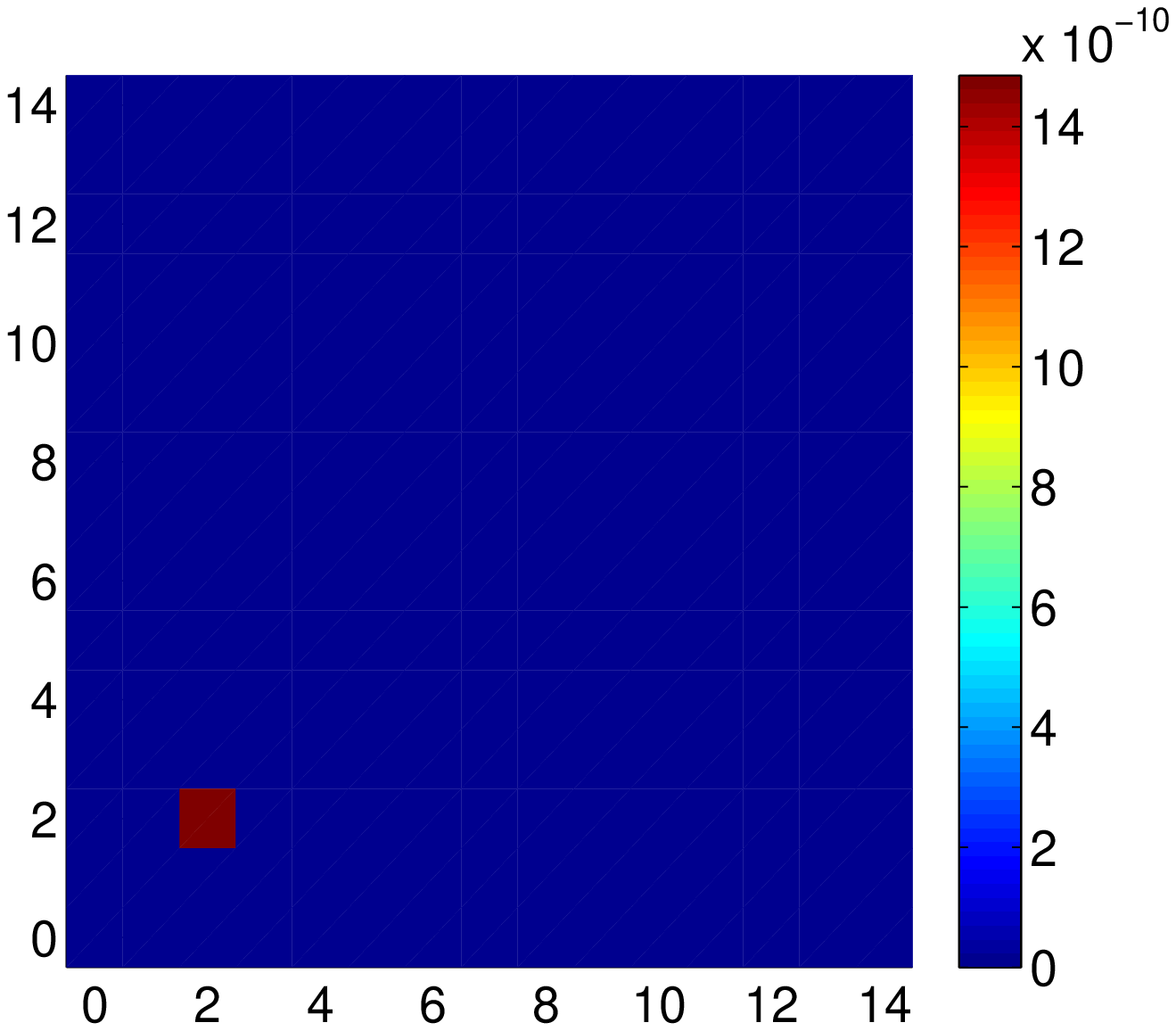}
\epsfxsize=5.35cm \epsfbox{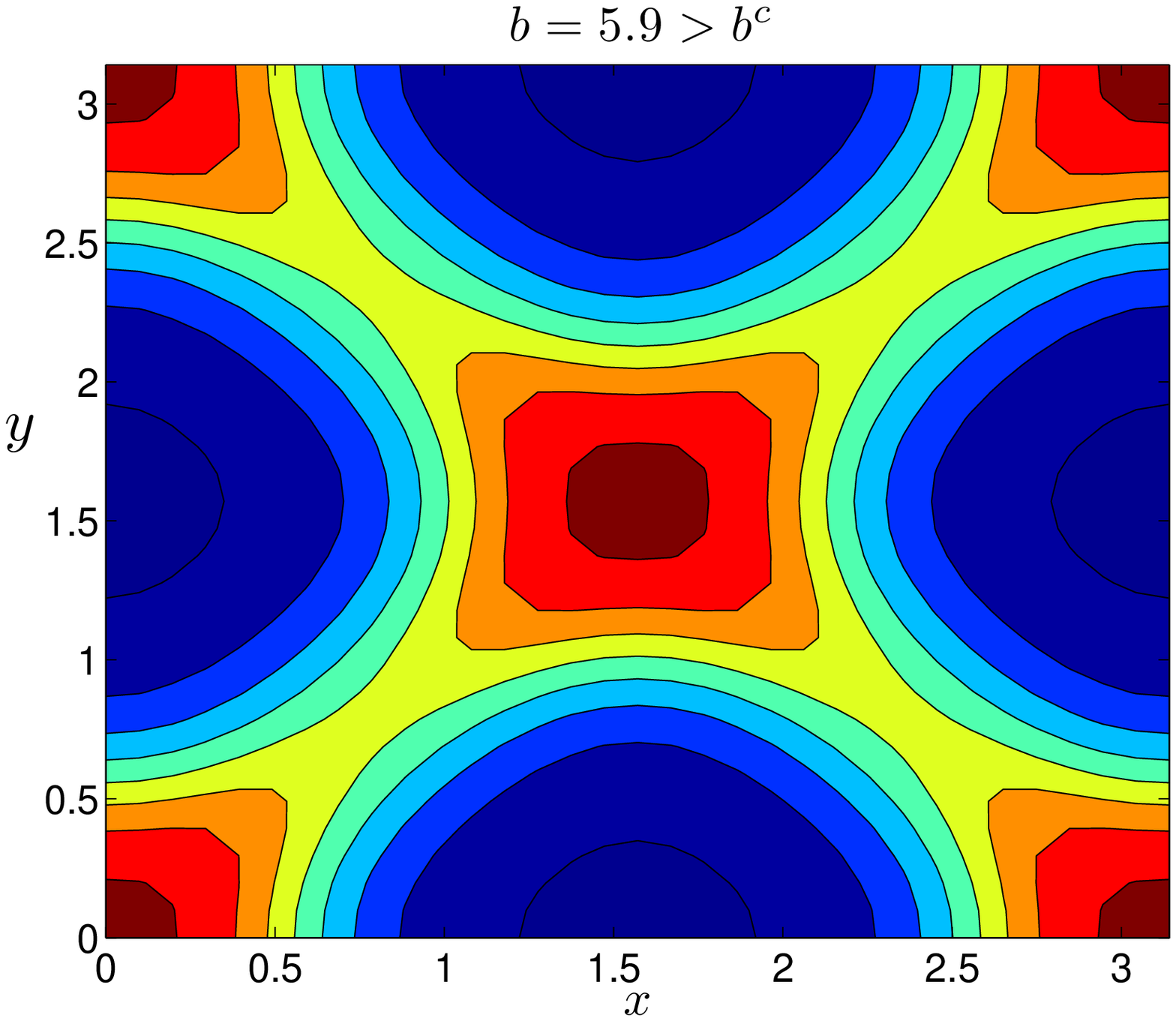} \epsfxsize=5.15cm\epsfbox{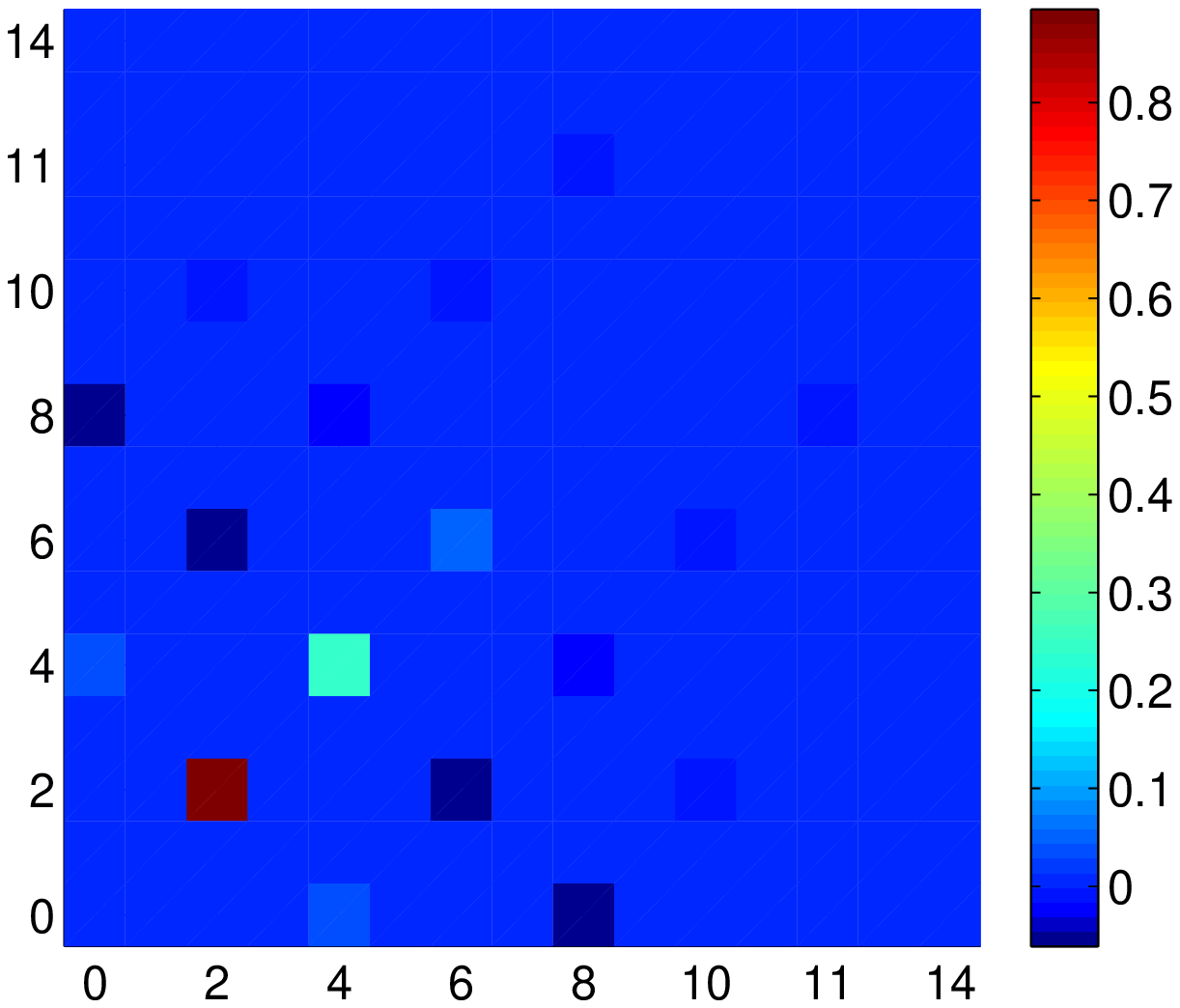}
\end{center}
\caption{Hysteresis cycle. The values of the parameters are the same as in Fig.~\ref{bif1d}. {\it Left.} Numerical solutions. {\it Right.} The spectrum of the solutions.}\label{hyst1amp}
\end{figure}

In Fig.~\ref{bif1d} we show the bifurcation diagram
for specific values of the parameters: the origin is locally
stable for $b <b^c$ and, when $b=b^c$, two backward-bending
branches of unstable fixed points bifurcate from the origin.
These unstable branches turn around and become stable at some
$b=b^s$ so that in the range $b^s<b<b^c$ two qualitatively
different stable states coexist, namely the origin and the large
amplitude branches. The existence of different stable states for
one single value of the parameter allows for the possibility of
hysteresis as $b$ is varied. In Fig. \ref{hyst1amp} we show  a
hysteresis cycle corresponding to a periodic variation of the
bifurcation parameter. Starting with a value of the parameter
above $b^c$ the solution jumps immediately to the stable branch
corresponding to a pattern whose amplitude is relatively
insensitive to the size of the bifurcation parameter. Decreasing
$b$ below the value $b^c$ the solution persists on the upper
branch and the pattern does not disappear. With a further decrease
of $b$ below $b^s$ the solution jumps to the constant steady
state. To have the pattern formation one has to increase the
parameter $b$ above $b^c$.

\begin{table}
\begin{center}
\caption{Subcritical case: Parameters as in Fig.\ref{compsub1amp}. A relatively good agreement is observed.}
\label{submode}
\vskip.1cm
\begin{tabular}{|ccc|}\hline
     Modes        & Numerical solution &  Approximated solution        \\ \hline
$\cos(2x)\cos(2y)$   &     0.0894     &    0.0909     \\ \hline
$\cos(4x)$    &    0.0420      &     0.0252     \\ \hline
$\cos(4y)$    &    0.0420      &  0.0252  \\ \hline
$\cos(4x)\cos(4y)$    & $0.2513$  &  $0.1418$ \\ \hline
\end{tabular}
\end{center}
\end{table}

\subsection{Double eigenvalue and secular terms at $O(\varepsilon^3)$}\label{mixedmode}

If the multiplicity in \eqref{4.6} is $m=2$ and the so called no-resonance condition
holds, namely:

\begin{eqnarray}
 \phi_i+\phi_j\neq \phi_j   \quad  &\mbox{or}&
\quad\psi_i-\psi_j\neq \psi_j \nonumber \\
 & \mbox{and}&  \label{4.13} \\
\phi_i-\phi_j\neq\phi_j \quad  &\mbox{or}& \quad
\psi_i+\psi_j\neq\psi_j   \nonumber
\end{eqnarray}
with $i,j=1,2$ and $i\neq j$,
then the Fredholm alternative is automatically satisfied at
$O(\varepsilon^2)$ by imposing $T_1=0$ and $b^{(1)}=0$ (as in the supercritical case
with $m=1$).
Secular terms appear into equation (\ref{4.10}) at
$O(\varepsilon^3)$, whose solvability condition leads to
the following two coupled Landau equations for the amplitudes $A_1$ and $A_2$:
\begin{subequations}\label{4.19}
\begin{eqnarray}\label{4.19a}
\frac{dA_1}{dT_2}&=&\sigma A_1-L_1 A_1^3+\Omega_1 A_1 A_2^2,\\
\frac{dA_2}{dT_2}&=&\sigma A_2-L_2 A_2^3+\Omega_2 A_1^2\,
A_2.\label{4.19b}
\end{eqnarray}
\end{subequations}
Therefore:

\textbf{WNL analysis results in the degenerate non resonant supercritical case}
\emph{
Assume that:
\begin{enumerate}
\item
$\ep^2=(b-b^c)/b^c$ is small enough so that
the uniform steady state $(u_0,v_0)$ in \eqref{equi_coe} is unstable
to modes corresponding only to the eigenvalue $\kcb$;
\item
there exists two couples of integers $(m_i,n_i),\,i=1,2$  such that:
$$
\kcb^2\equiv \phi^2_i+\psi^2_i\ {\rm where}\ \ \phi_i\equiv \frac{m_i\pi}{L_x},\ \
\psi_i\equiv  \frac{n_i\pi}{L_y}\, ;
$$
\item
$\phi_i$ and $\psi_i$ satisfy the no-resonance condition \eqref{4.13};
\item
(supercriticality) the system \eqref{4.19} admits at least one stable equilibrium.
\end{enumerate}
Then the emerging asymptotic solution of the reaction-diffusion system \eqref{1.1} at the leading order is approximated by:
$$
\mathbf{w}=\varepsilon \bfrho (A_{1\infty}\cos(\phi_1 x)\cos(\psi_1 y)+A_{2\infty}\cos(\phi_2 x)\cos(\psi_2 y))+O(\varepsilon^2),
$$
where $(A_{1\infty}, A_{2\infty})$ is a stable stationary state of the system \eqref{4.19}.
}

The stationary solutions of the equations
(\ref{4.19a})-(\ref{4.19b}) are given by the trivial
equilibrium and the following eight points:
\begin{equation}\label{4.28}
P_1^{\pm}\equiv\left(\pm \sqrt{\frac{\sigma}{L_1}},0\right);\qquad
P_2^{\pm}\equiv\left(0, \pm \sqrt{\frac{\sigma}{L_2}}\,\right);
\end{equation}
\begin{equation}\label{4.29}
P_3^{(\pm, \pm)}\equiv\left(\pm\sqrt{\frac{\sigma(L_2+\Omega_1)}{L_1L_2-\Omega_1\Omega_2}},
\pm\sqrt{\frac{\sigma(L_1+\Omega_2)}{L_1L_2-\Omega_1\Omega_2}}\,\right)\,
,
\end{equation}
where the first coordinate is the amplitude $A_1$ and the second one is $A_2$.
It is straightforward to prove that the trivial equilibrium is always unstable.

The results of the linear stability analysis of the equilibrium points are
summarized  in the following table:
\vskip.5cm
\begin{center}
\begin{tabular}{ccc}
  \hline
  $\ $ & Existence & Stability \\
  \hline
  \hline\\
  $P_1^{\pm}$ & $L_1>0$ & $L_1+\Omega_{2}<0$ \\
  \ \\
  \hline\\
  $P_2^{\pm}$ & $L_2>0$ & $L_2+\Omega_{1}<0$ \\
   \ \\
  \hline\\
  $\ $ &
  $
  \left\{
  \begin{array}{lll}
  L_1L_2-\Omega_1\Omega_2<0,\\
  L_1+\Omega_2<0\\
  L_2+\Omega_1<0
  \end{array}
  \right.
  $
  & always unstable \\
  $P_3^{(\pm, \pm)}$ & or & $\ $ \\
  $\ $ &
  $
  \left\{
  \begin{array}{lll}
  L_1L_2-\Omega_1\Omega_2>0,\\
  L_1+\Omega_2>0\\
  L_2+\Omega_1>0
  \end{array}
  \right.
  $
  & $L_1\Omega_1+L_2\Omega_2+2L_1L_2<0$ \\
  \ \\
  \hline\hline
\end{tabular}
\end{center}
\vskip.5cm
%
%
%
%

%
%
%
%

From the above table it can be easily seen that when $P_3^{(\pm, \pm)}$ exist stable, the equilibria $P_j^{\pm}$, with $j=1,2$ are unstable.
Therefore,  when pattern forms, there are two possible asymptotic behaviors of the solution:
a mixed mode steady state solution arising in correspondence of the stable equilibria
$P_3^{(\pm, \pm)}$ and a single mode steady state solution when $P_j^{\pm}$, with $j=1,2$ are stable.
We also notice that, for a square domain, there is symmetry of the modes
(i.e. $\phi_1=\psi_2$ and $\phi_2=\psi_1$), and one always has $L_1=L_2$ and $\Omega_1=\Omega_2$
(this is obvious for symmetry reasons, and can also be seen by inspection of the formulas
\eqref{l2d}-\eqref{o2d});
which implies that the conditions for the existence and stability of $P_1^{\pm}$ and $P_2^{\pm}$ coincide.

In what follows we shall perform one numerical test concerning single mode patterns, and
three tests concerning the case of mixed modes patterns.

\begin{figure}
\begin{center}
\epsfxsize=5.65cm \epsfbox{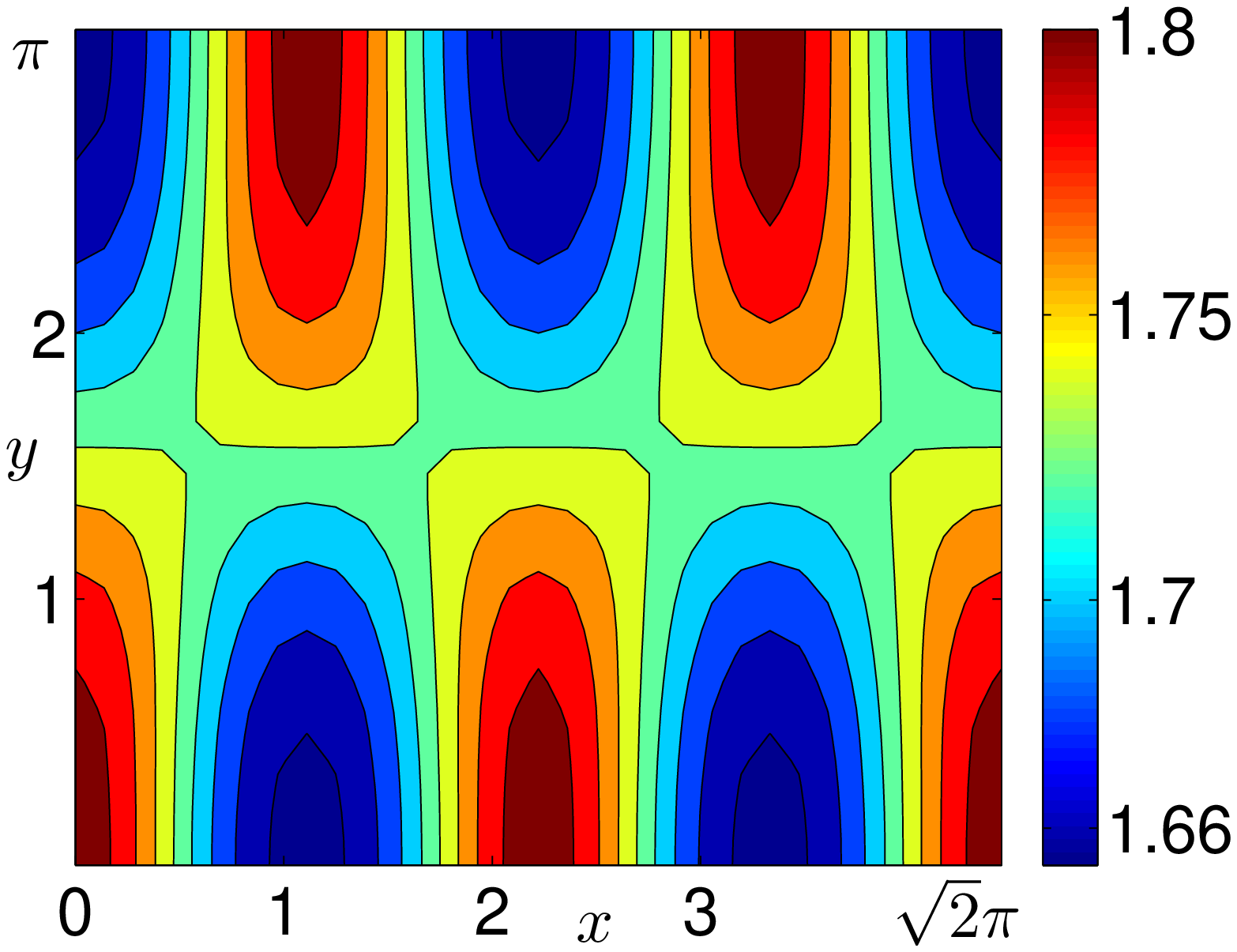} \epsfxsize=5.77cm
\epsfbox{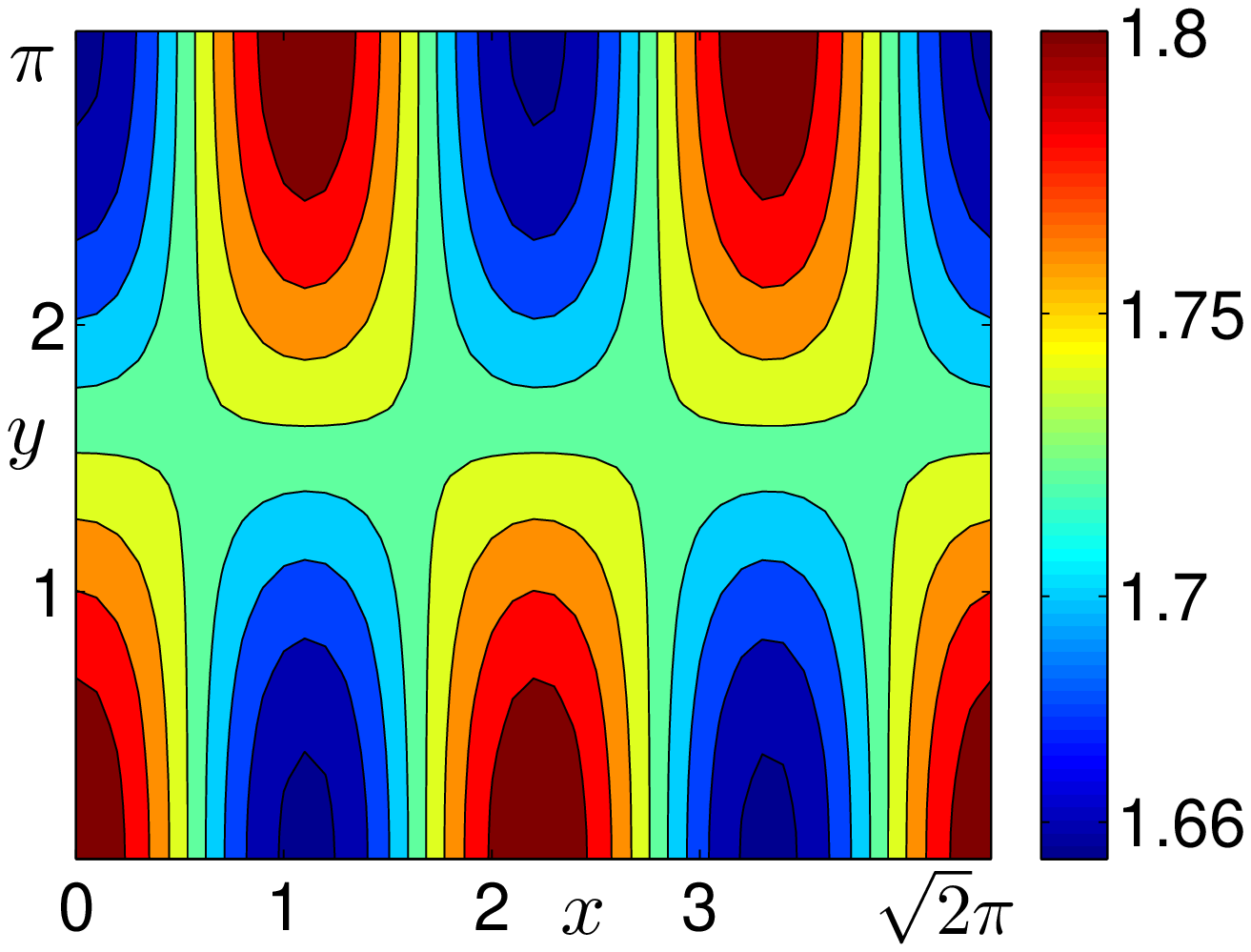}
\end{center}
\caption{Comparison between the numerical solution (on the left)
and the weakly nonlinear first order approximation of the solution
(on the right). The system parameters are chosen as follows:
$\Gamma=52.453$, $\mu_1=1.2$, $\mu_2=1$, $\gamma_{11}=0.5$,
$\gamma_{12}=0.4$, $\gamma_{21}=0.38$, $\gamma_{22}=0.41$,
$c_1=c_2=0.2$, $a_1=0.01$, $b^c=7.316$,
$\varepsilon=0.1$, $b=(1+0.1^2)b^c=7.389$, $a_2=0.1$, $b_2=0.7$.}\label{9_n}
\end{figure}

In our first test we consider the rectangular domain where
$L_x=\sqrt{2}\pi$ and $L_y=\pi$ and with the choice of the parameters as in the
caption of Fig.\ref{9_n}, the unique discrete unstable mode is $\kcb^2=9$.
The mode pairs satisfying the condition \eqref{kcbuguamn} are $(0,3)$ and $(4, 1)$.
Moreover the equilibria $P_2^{\pm}$ and $P_3^{(\pm, \pm)}$ are unstable and only
the steady states $P_1^{\pm}$ are stable.
The predicted asymptotic solution therefore is:
\begin{equation}\label{ex9}
\textbf{w}=\varepsilon A_{1\infty}\bfrho
\cos\left(2\sqrt{2}x\right)\cos(y)+O(\varepsilon^2),
\end{equation}
where $A_{1\infty}$ is the nonzero coordinate of one of the points $P_1^{\pm}$
(which equilibrium is reached depends on initial conditions).
Our numerical tests starting from a random periodic
perturbation of the equilibrium show that the solution evolves to the rectangular
pattern predicted in \eqref{ex9}. Figure \ref{9_n} shows the agreement (with $\varepsilon=0.1$)
between the numerical solution and the solution expected on the basis of the weakly nonlinear analysis.

\begin{figure}
\begin{center}
\epsfxsize=5.65cm \epsfbox{vuoto.eps} \epsfxsize=5.65cm
\epsfbox{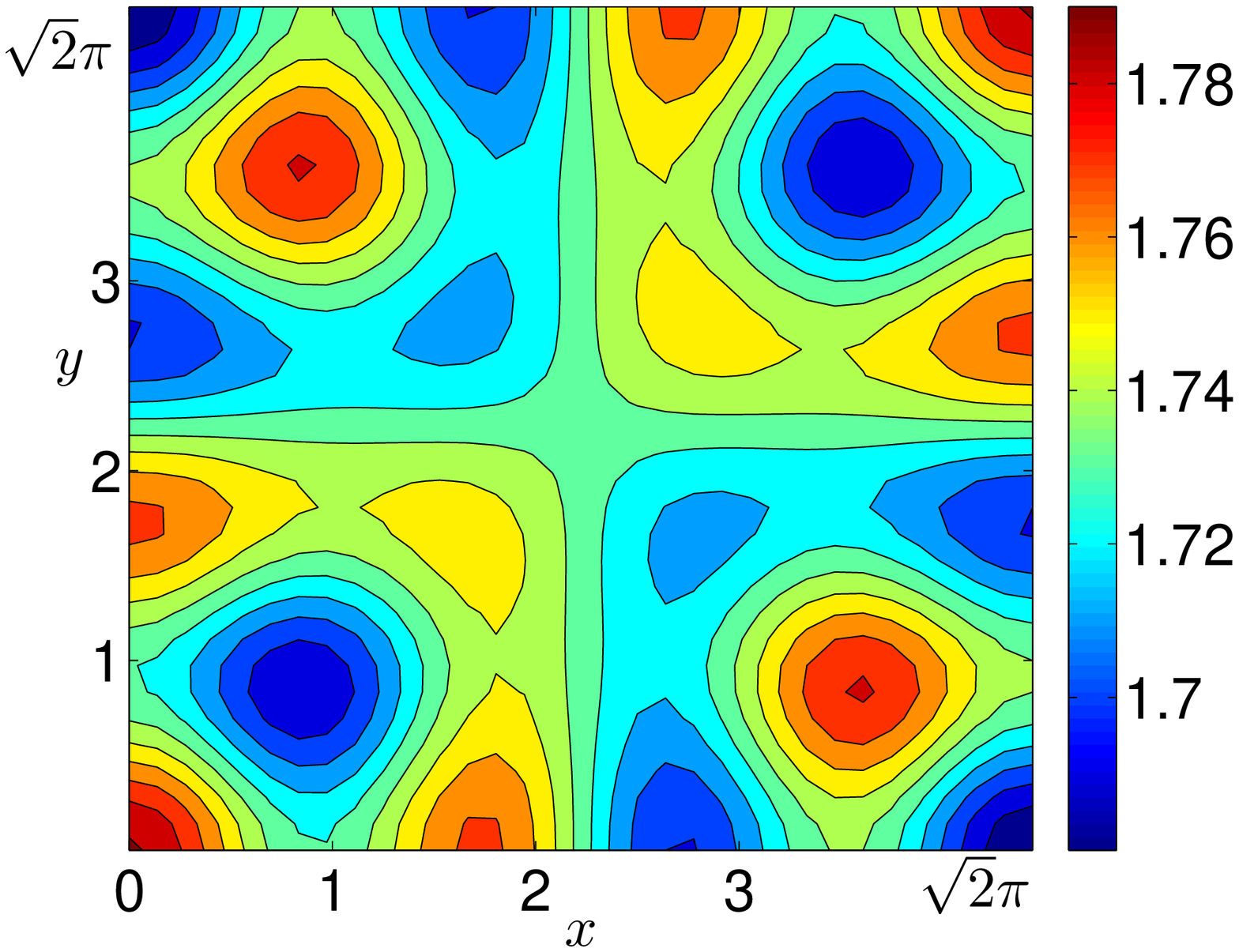}
\end{center}
\caption{Comparison between the numerical solution (on the left)
and the weakly nonlinear first order approximation of the solution
(on the right). Regions where $u \geq 1.735$ are shaded in both
plots. The system parameters are chosen as follows:
$\Gamma=49.95$, $\mu_1=1.2$, $\mu_2=1$, $\gamma_{11}=0.5$,
$\gamma_{12}=0.4$, $\gamma_{21}=0.38$, $\gamma_{22}=0.41$,
$c_1=c_2=0.2$, $a_1=0.1$, $a_2=0.015$, $b^c=4.354$, $\varepsilon=0.02$,
$b=(1+0.02^2)b^c=4.356$, $b_2=0.2$. }\label{mixed2}
\end{figure}
\begin{figure}
\begin{center}
\epsfxsize=5.65cm \epsfbox{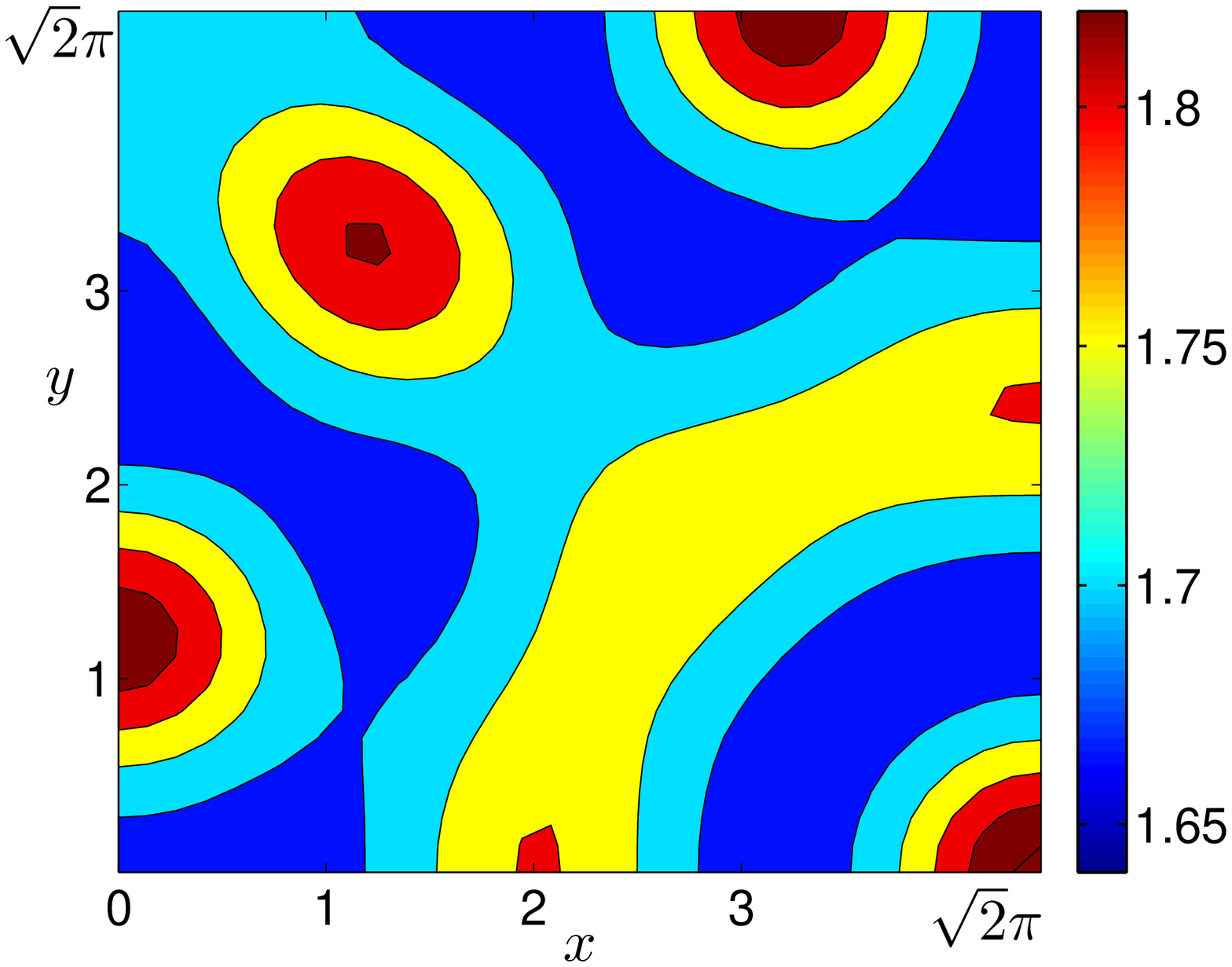} \epsfxsize=5.65cm
\epsfbox{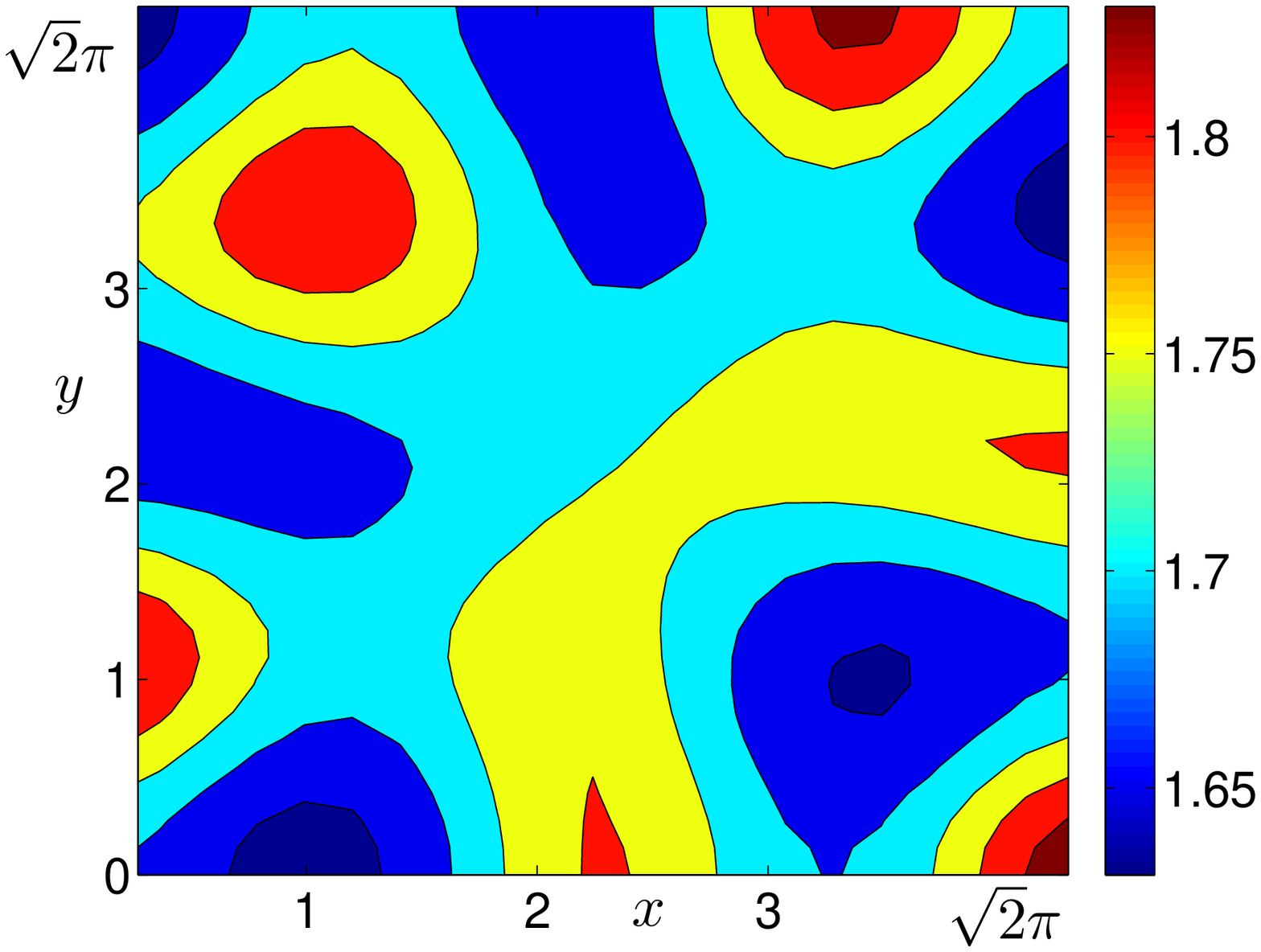}
\end{center}
\caption{Comparison between the numerical solution (on the left)
and the weakly nonlinear first order approximation of the solution
(on the right). Regions where $u \geq 1.735$ are shaded in both
plots. The system parameters are chosen as follows:
$\Gamma=24.517$, $\mu_1=1.2$, $\mu_2=1$, $\gamma_{11}=0.5$,
$\gamma_{12}=0.4$, $\gamma_{21}=0.38$, $\gamma_{22}=0.41$,
$c_1=c_2=0.2$, $a_1=a_2=0.01$, $b^c=3.024$, $\varepsilon=0.1$,
$b=(1+0.1^2)b^c=3.054$, $b_2=0.1$. }\label{8_5}
\end{figure}

In the second numerical test we consider a square domain with dimensions $L_x=L_y=\sqrt{2}\pi$
and choose the parameter values in such a way that only the most unstable
mode $\kcb^2=13$ falls within the band of unstable modes.
The set of parameters is described in the caption of Fig.\ref{8_5}.
The uniform steady state is then linearly unstable to the two mode pairs
$(1,5)$ and $(5,1)$. With this choice of the parameters the two
single mode steady states in $\eqref{4.28}$ are unstable and the
mixed mode steady states in $\eqref{4.29}$ are stable.
Therefore the predicted equilibrium solution to first order is:
\begin{equation}
\textbf{w}=\varepsilon \left(A_{1\infty}\bfrho
\cos\left(\frac{x}{\sqrt{2}}\right)\cos\left(\frac{5y}{\sqrt{2}}\right)+A_{2\infty}\bfrho
\cos\left(\frac{5x}{\sqrt{2}}\right)\cos\left(\frac{y}{\sqrt{2}}\right)\right)+O(\varepsilon^2),
\end{equation}
where $A_{1\infty}, A_{2\infty}$ are the coordinates of one of the equilibrium points in
$\eqref{4.29}$. Which of these equilibrium points is reached clearly depends on the initial conditions.
In Fig.\ref{mixed2} the comparison between the numerical
simulation of the original system and the expected solution with
$\varepsilon=0.02$ shows a good agreement and the values of the most excited modes of the numerical solution and of the solution computed
using the weakly nonlinear analysis are quite similar, respectively 0.0278 and 0.0282.

In the third numerical test we consider again the square domain $L_x=L_y=\sqrt{2}\pi$
and the parameter values are chosen as in the caption of Fig.\ref{8_5}, in such a way that only the most unstable
mode $\kcb^2=8.5$ falls within the band of unstable modes. In this domain
the uniform steady state is then linearly unstable to the two mode pairs
$(1,4)$ and $(4,1)$. As in the previous test, the only stable states are the
mixed mode steady states in $\eqref{4.29}$ and the predicted equilibrium solution, truncated at the first order, is:
\begin{equation}
\textbf{w}=\varepsilon \left(A_{1\infty}\bfrho
\cos\left(\frac{x}{\sqrt{2}}\right)\cos\left(\frac{4y}{\sqrt{2}}\right)+A_{2\infty}\bfrho
\cos\left(\frac{4x}{\sqrt{2}}\right)\cos\left(\frac{y}{\sqrt{2}}\right)\right)+O(\varepsilon^2),
\end{equation}
where $A_{1\infty}, A_{2\infty}$ are the coordinates of one of the equilibrium points in
$\eqref{4.29}$.
In Fig.\ref{8_5} we show the comparison between the numerical
simulation of the original system and the expected solution with
$\varepsilon=0.1$. The two solutions are qualitatively similar, in
particular the numerical solution evolves to a mixed mode steady
state.
However one can see a significant quantitative discrepancy.
A closer look, see Table \ref{tabfig33}, reveals that this discrepancy is due to the presence of the
subharmonic $(3,3)$ which is of the same order of magnitude of the modes $(1,4)$ and $(4,1)$,
while the weakly nonlinear analysis predicts to be $O(\varepsilon^2)$ (see formula \eqref{sol_ris}).
It is interesting that the same discrepancy was found
in a numerical test performed in \cite{CMM97} for a different type of reaction-diffusion system,
where the uniform steady state was linearly
unstable to the same two mode pairs $(1,4)$ and $(4,1)$.
Noticing that the subharmonic $(3,3)$ corresponds to the
discrete eigenvalue $k^2=9$ (according to the formula \eqref{con2d}),
one might conjecture that the closeness to $\kcb^2=8.5$ makes the linear damping mechanism
unable to overcome the quadratic excitation mechanism coming from the interaction of the main
harmonics $(1,4)$ and $(4,1)$.

\begin{figure}
\begin{center}
\epsfxsize=5.65cm \epsfbox{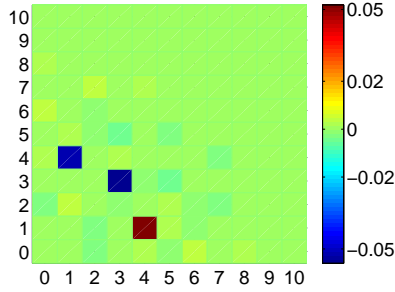}
\end{center}
\caption{The spectrum of the numerical solution on the left of Fig. \ref{8_5}.}\label{8_5spectrum}
\end{figure}
\begin{table}
\begin{center}
\caption{}
\label{tabfig33}
\vskip.1cm
\begin{tabular}{|ccc|}\hline
     Modes        & Numerical solution &  Approximated solution        \\ \hline
     $\cos(x/\sqrt{2})\cos(4y/\sqrt{2})$   &     0.0519     &    0.0688       \\ \hline
$\cos(4x/\sqrt{2})\cos(y/\sqrt{2})$    &    0.0519      &     0.0688     \\ \hline
$\cos(3x/\sqrt{2})\cos(3y/\sqrt{2})$    &    0.0560      &  0.0092  \\ \hline
\end{tabular}
\end{center}
\end{table}

\begin{figure}
\begin{center}
\epsfxsize=5.65cm \epsfbox{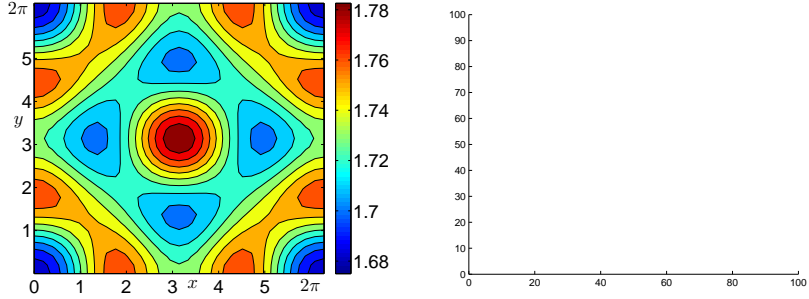} \epsfxsize=5.65cm
\epsfbox{vuoto.eps}
\end{center}
\caption{Comparison between the numerical solution (on the left)
and the weakly nonlinear first order approximation of the solution
(on the right). Regions where $u \geq 1.735$ are shaded in both
plots. The system parameters are chosen as follows:
$\Gamma=30.6$, $\mu_1=1.2$, $\mu_2=1$, $\gamma_{11}=0.5$,
$\gamma_{12}=0.4$, $\gamma_{21}=0.38$, $\gamma_{22}=0.41$,
$c_1=c_2=0.2$, $a_1=a_2=0.1$, $b_2=0.5$, $b^c=7.181$, $\varepsilon=0.05$,
$b=7.199$. }\label{supsq}
\end{figure}

In our final test we consider the case of the supersquare as discussed in the paper \cite{DSS97}, characterized
by having $\kcb^2=5$ on a square domain.
We consider a domain with dimensions $L_x=L_y={2}\pi$
and choose the parameter values in such a way that only the most unstable discrete
mode $\bar{k}_c^2=5$ falls within the band of unstable modes allowed
by the boundary conditions.
The uniform steady state is then linearly unstable to the two mode pairs
$(1,2)$ and $(2,1)$. With this choice of the parameters the two
single mode steady states in $\eqref{4.28}$ are unstable and the
mixed mode steady states in $\eqref{4.29}$ are stable.
Notice that we shall consider only the positive equilibrium $P_3^{(+,+)}$.
Therefore the predicted solution truncated at the first order is:
\begin{equation}
\textbf{w}=\varepsilon \left(A_{1\infty}\bfrho
\cos\left(x\right)\cos\left(2y\right)+A_{2\infty}\bfrho
\cos\left(2x\right)\cos\left(y\right)\right)+O(\varepsilon^2),
\end{equation}
where $A_{1\infty}, A_{2\infty}$ are the positive values in $\eqref{4.29}$. In
Fig.\eqref{supsq} we show the comparison between the numerical
simulation of the original system and the solution predicted by the weakly nonlinear
analysis with $\varepsilon=0.05$.
The two solutions are very close and one can see that the accuracy is,
as expected, $O(\ep^2)$.

\subsection{Double eigenvalue and secular terms at $O(\varepsilon^2)$}

If the multiplicity of the eigenvalue is $m=2$ and the following resonance
condition is satisfied:
\begin{eqnarray}
 \phi_i+\phi_j = \phi_j   \quad  &\mbox{and}&
\quad\psi_i-\psi_j = \psi_j \nonumber \\
 & \mbox{or}&  \label{res_si} \\
\phi_i-\phi_j  =\phi_j \quad  &\mbox{and}& \quad
\psi_i+\psi_j = \psi_j   \nonumber
\end{eqnarray}
 with $i, j=1, 2$  and $i\neq j$,
then secular terms appear at $O(\varepsilon^2)$ in \eqref{4.7}.
In what follows, without loss of generality, we shall thus perform the weakly nonlinear analysis imposing
that the second condition in \eqref{res_si} hold, with $i=2$ and $j=1$.
Taking into account these conditions
and the relation \eqref{kcbuguamn}, it follows that $\phi_2=2\phi_1$, $\psi_2=0$,
$\psi_1= \sqrt{3}\phi_1$ and $\phi_1=  \kcb/2$, see \eqref{esa_vec}.
Hexagonal patterns fall within this class of solutions, as shown in \cite{CMM97}.
We also notice that the above relations imply that $L_y=\sqrt{3}L_x$.

The solution \eqref{4.6} at the first order reads:
\begin{equation}\label{4.15}
{\bf w}_1=A_1(T_1, T_2)\bfrho\cos(\phi_1 x)\cos(\sqrt{3}\phi_1 y)+
A_2(T_1, T_2)\bfrho\cos(2\phi_1 x).
\end{equation}
The solvability condition at
$O(\varepsilon^2)$ gives the following system
of equations for the two amplitudes $A_1, A_2$ :
\begin{subequations}\label{4.16}
\begin{eqnarray}\label{4.16a}
\frac{\partial A_1}{\partial T_1}&=&\sigma A_1-L A_1A_2,\\
\frac{\partial A_2}{\partial T_1}&=&\sigma A_2-\frac{L}{4}\ A_1^2\, .\label{4.16b}
\end{eqnarray}
\end{subequations}
%

The stationary solutions of the equations
\eqref{4.16} are the trivial equilibrium and $Q^{\pm}\equiv\left(\pm\ {2\sigma}/{L}, {\sigma}/{L}\right)$.
It is easy to see that the nontrivial stationary solutions associated with $Q^{\pm}$
are always unstable (the eigenvalues of the jacobian matrix evaluated at $Q^{\pm}$
are $\lambda_1=-\sigma, \lambda_2=2\sigma$); therefore the weakly
nonlinear analysis, at this order, is not able to predict the amplitude of the
pattern.
This is a subcritical bifurcation case and the asymptotic analysis has to be pushed to
higher order in the amplitude to obtain qualitatively reliable results \cite{BMS09}.
To $O(\varepsilon^3)$ one finds the following system
for the amplitudes $A_1$ and $A_2$:
\begin{subequations}\label{4.32}
\begin{eqnarray}\label{4.32a}
\frac{dA_1}{dT}&=&\bar{\sigma}_1 A_1-\bar{L}_1 A_1A_2+\bar{\alpha}_1A_1^3+\bar{\beta}_1 A_1 A_2^2,\\
\frac{dA_2}{dT}&=&\bar{\sigma}_2 A_2-\bar{L}_2
A_1^2+\bar{\alpha}_2A_2^3+\bar{\beta}_2 A_1^2\, A_2.\label{4.32b}
\end{eqnarray}
\end{subequations}
%

\textbf{WNL analysis results in the degenerate resonant case}
\emph{
Assume that:
\begin{enumerate}
\item
$\ep=(b-b^c)/b^c$ is small enough so that
the uniform steady state $(u_0,v_0)$ in \eqref{equi_coe} is unstable
to modes corresponding only to the eigenvalue $\kcb$;
\item
there exists two couples of integers $(m_i,n_i),\,i=1,2$  such that:
$$
\kcb^2\equiv \phi^2_i+\psi^2_i\ {\rm where}\ \ \phi_i\equiv \frac{m_i\pi}{L_x},\ \
\psi_i\equiv  \frac{n_i\pi}{L_y} \, ;
$$
\item
$\phi_i$ and $\psi_i$ satisfy the resonance condition \eqref{res_si};
\item
the system \eqref{4.32} admits at least one stable equilibrium.
\end{enumerate}
Then the emerging asymptotic solution of the reaction-diffusion system \eqref{1.1} at the leading order is approximated by:
$$
\mathbf{w}=\varepsilon \bfrho (A_{1\infty}\cos(\phi_1 x)\cos(\psi_1 y)+A_{2\infty}\cos(\phi_2 x)\cos(\psi_2 y))+O(\varepsilon^2),
$$
where $(A_{1\infty}, A_{2\infty})$ is a stable stationary state of the system \eqref{4.32}.
These solutions are rolls (when $A_{1\infty}=0$)  or hexagons.
}

%
%
%
%
%
%
%
%

Other than the trivial one, the equilibria of the equations \eqref{4.32} are
the points $R^{\pm}\equiv(0, \pm\sqrt{-{\bar{\sigma}_2}/{\bar{\alpha}_2}})$
which (when exist real) correspond to rolls,
and the six roots $H^{\pm}_i\equiv(A_{1i}^\pm, A_{2i}),\ i=1,2,3,$ of the following system:
\begin{equation}\label{4.35}
\left\{
\begin{split}
A_2^3\left(\bar{\alpha}_1\bar{\alpha}_2-\bar{\beta}_1\bar{\beta}_2\right)+&A_2^2\left(\bar{L}_1\bar{\beta}_2+
\bar{L}_2\bar{\beta}_1\right)+\\
&A_2\left(\bar{\alpha}_1\bar{\sigma}_2-\bar{L}_1\bar{L}_2-\bar{\beta}_2\bar{\sigma}_1\right)+\bar{L}_2\bar{\sigma}_1=0,\\
A_1^2=\displaystyle\frac{1}{\bar{\alpha}_1}\left(-\bar{\beta}_1A_2^2\right.+&\left.\bar{L}_1A_2-\bar{\sigma}_1 \right).
\end{split}
\right.
\end{equation}
The roots $H^{\pm}_i$, when exist real, correspond to hexagons.
The equilibria $R^{\pm}$ exist stable when $\bar{\alpha}_2<0$ and
 $\bar{L}_1>\sqrt{-\bar{\alpha}_2/\bar{\sigma}_2}(\bar{\sigma}_1-\bar{\beta}_1\bar{\sigma}_2/\bar{\alpha}_2)$.
When these points are stable the system \eqref{1.1}  support rolls of the form:
\begin{equation}
{\bf w}=\varepsilon A_{2\infty}\bfrho\cos(2\phi_1 x)+O(\varepsilon^2),
\end{equation}
where $A_{2\infty}$ is the nonzero coordinate of $R^{\pm}$.

\begin{figure}
\begin{center}
\epsfxsize=5.25cm \epsfbox{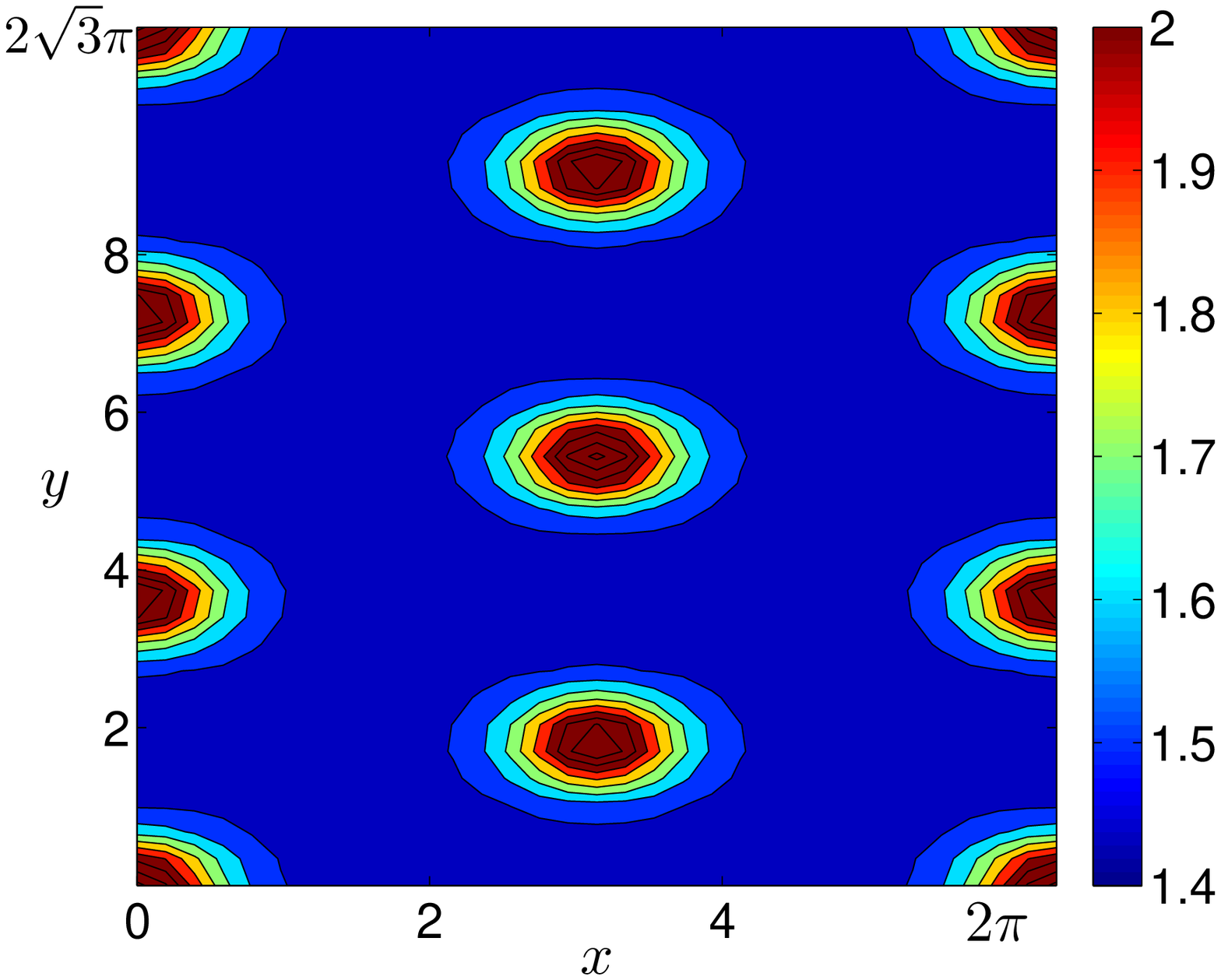} \epsfxsize=5.35cm
\epsfbox{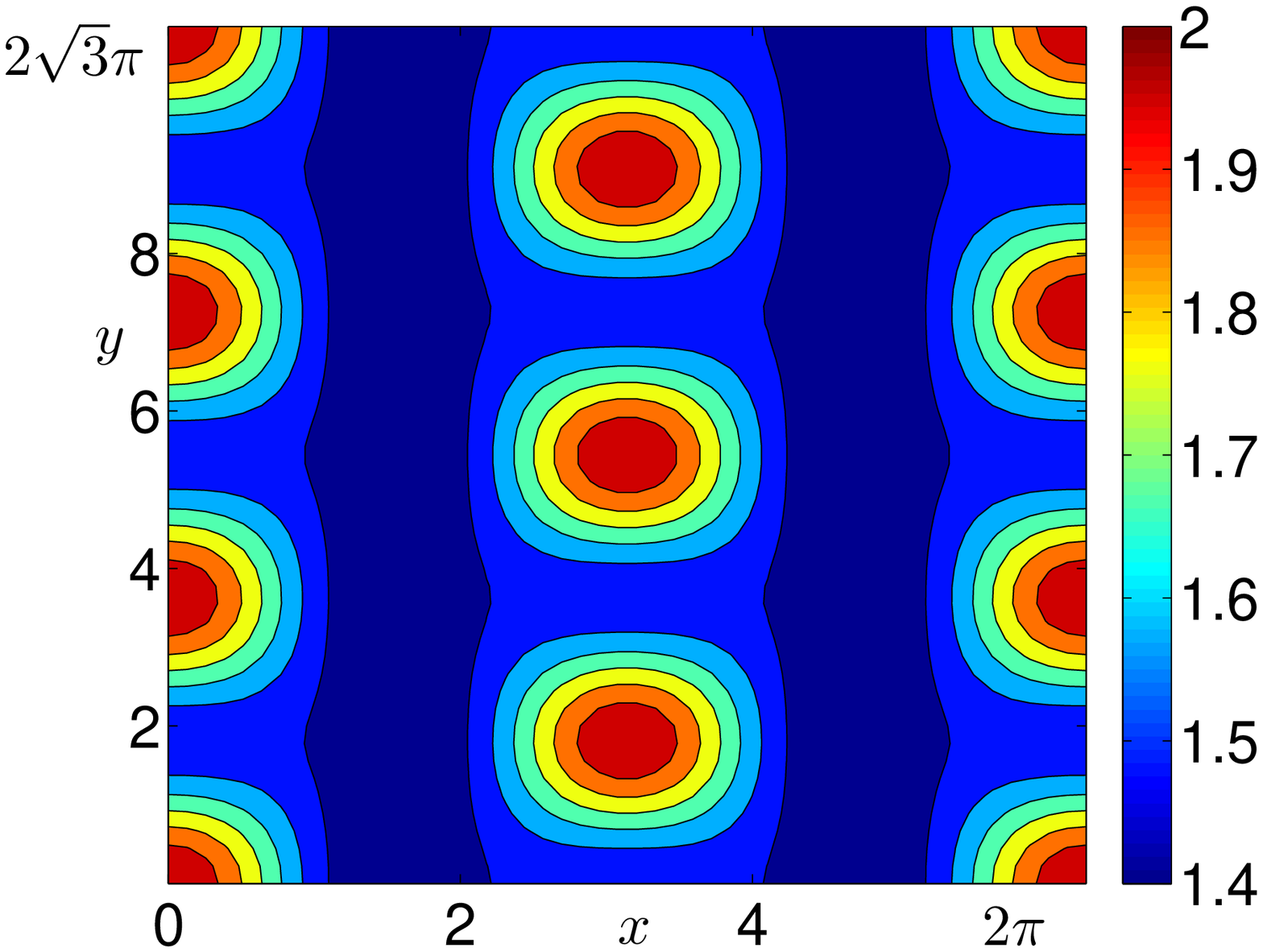}
\end{center}
\caption{The numerical solution (on the left) shows hexagonal
patterns. The expected solution via weakly nonlinear analysis up to the $O(\varepsilon^2)$ is
shown on the right. The system parameter are chosen:
$\Gamma=13.6$, $\mu_1=1.2$, $\mu_2=1$, $\gamma_{11}=0.5$,
$\gamma_{12}=0.4$, $\gamma_{21}=0.38$, $\gamma_{22}=0.39$,
$c_1=c_2=0.2$, $a_1=0.01$, $a_2=0.001$, $b_2=0.031$, $b^c=3.599$, $\ep=0.006$ and $b=3.62$. }\label{esagoni}
\end{figure}

\begin{figure}
\begin{center}
\epsfxsize=7.65cm \epsfbox{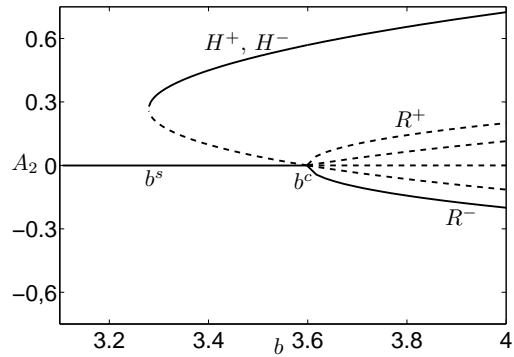}
\end{center}
\caption{The bifurcation diagram of the Stuart-Landau system \eqref{4.32}. The parameters are chosen as in Fig.\ref{esagoni}.}\label{bif_RH}
\end{figure}

\begin{figure}
\begin{center}
\epsfxsize=6.5cm \epsfbox{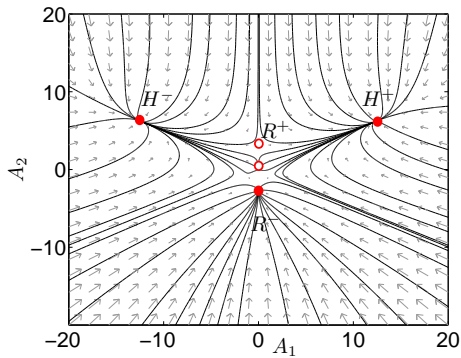}
\end{center}
\caption{Basins of attraction of rolls and hexagons. The parameters are chosen as in Fig.\ref{esagoni}.}\label{bas_RH}
\end{figure}

\begin{figure}
\epsfxsize=5.45cm \epsfbox{vuoto.eps} \epsfxsize=5.55cm
\epsfbox{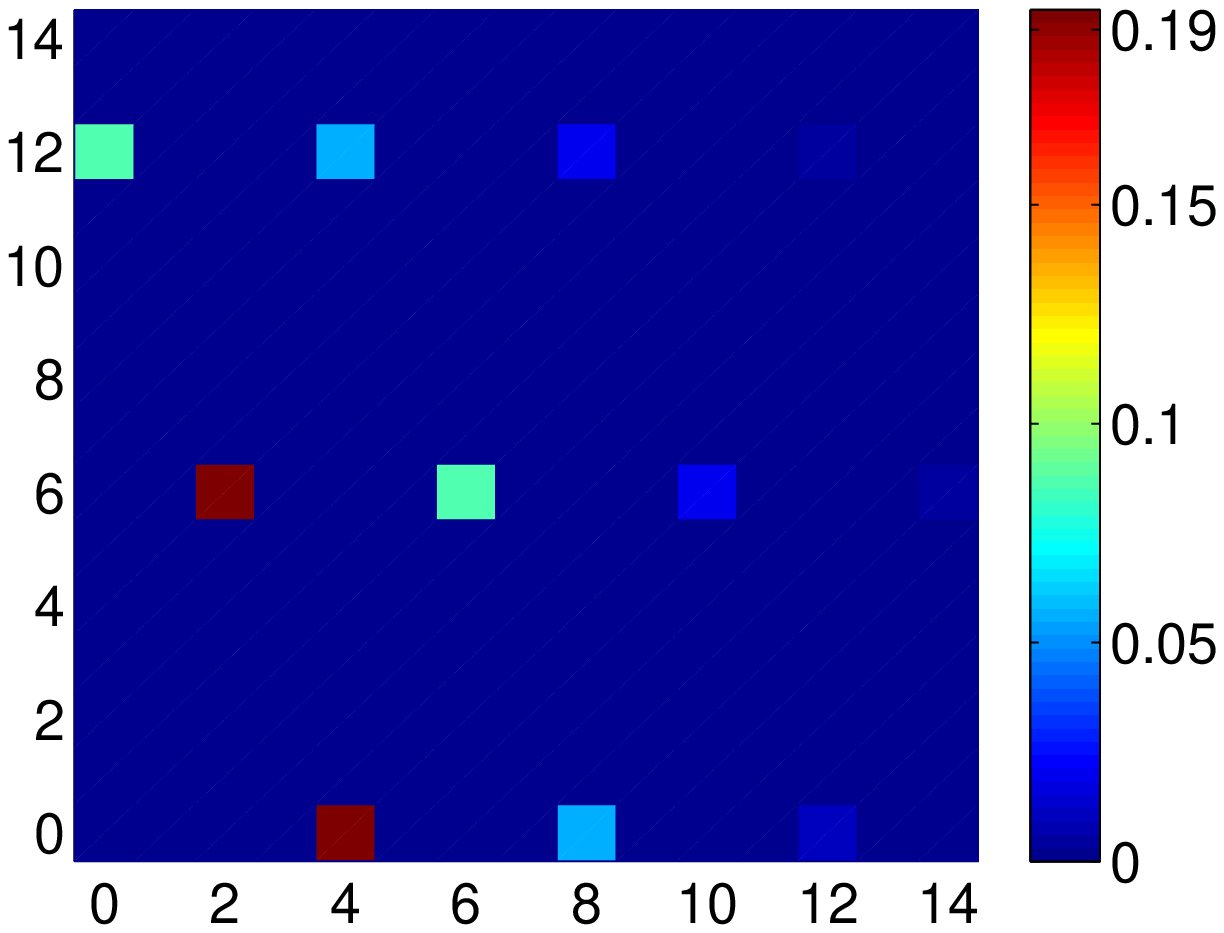}
\epsfxsize=5.45cm \epsfbox{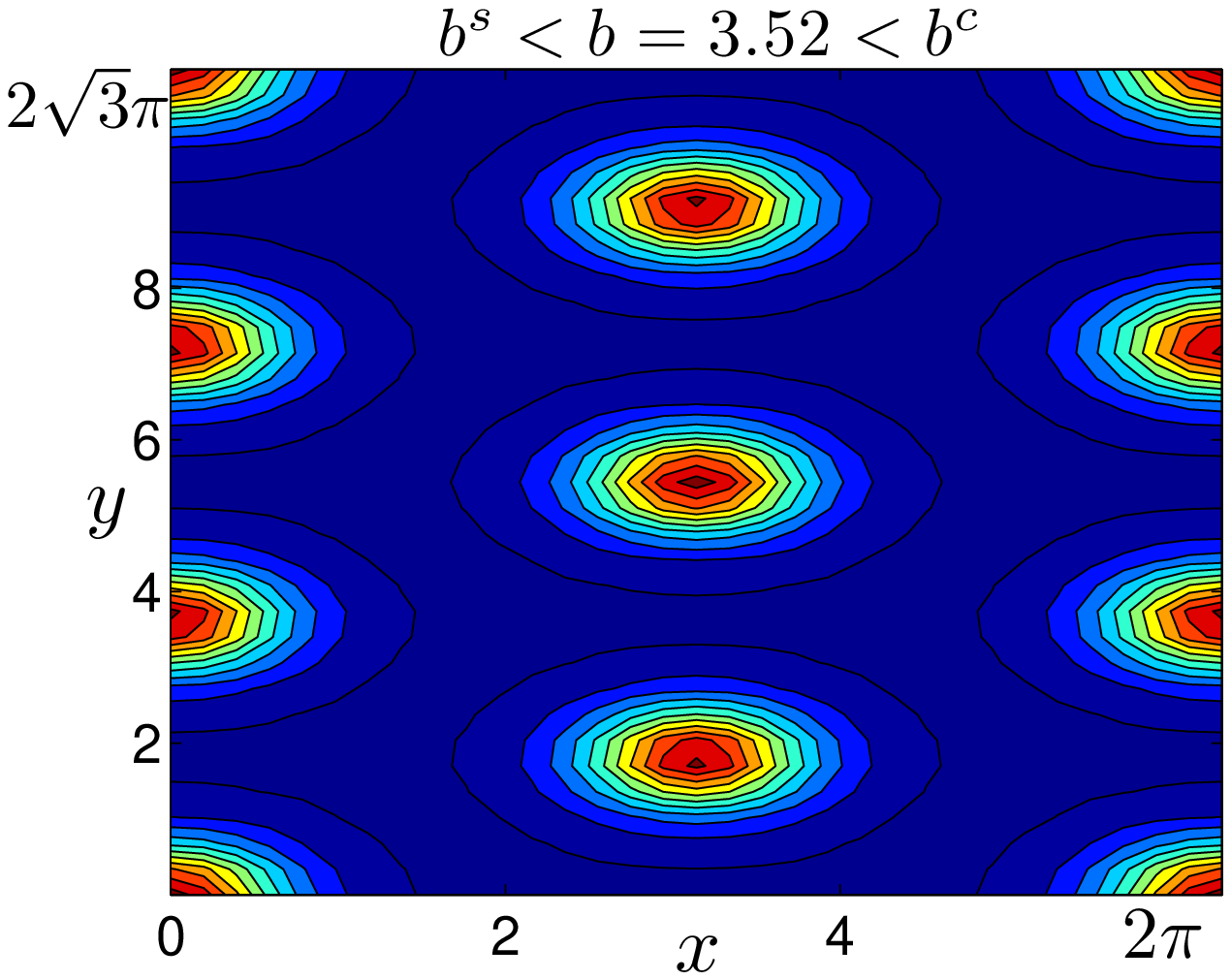} \epsfxsize=5.65cm
\epsfbox{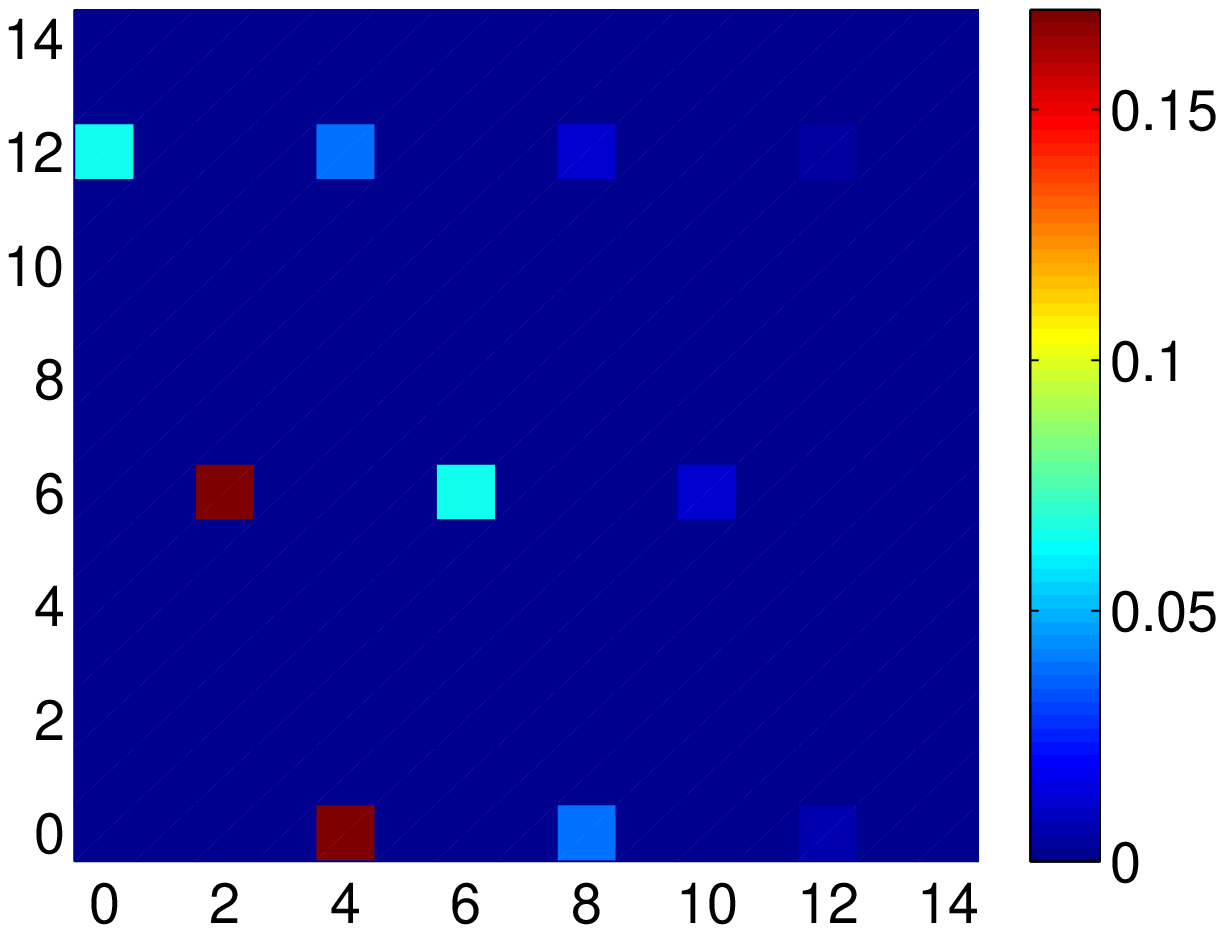}
\epsfxsize=5.45cm \epsfbox{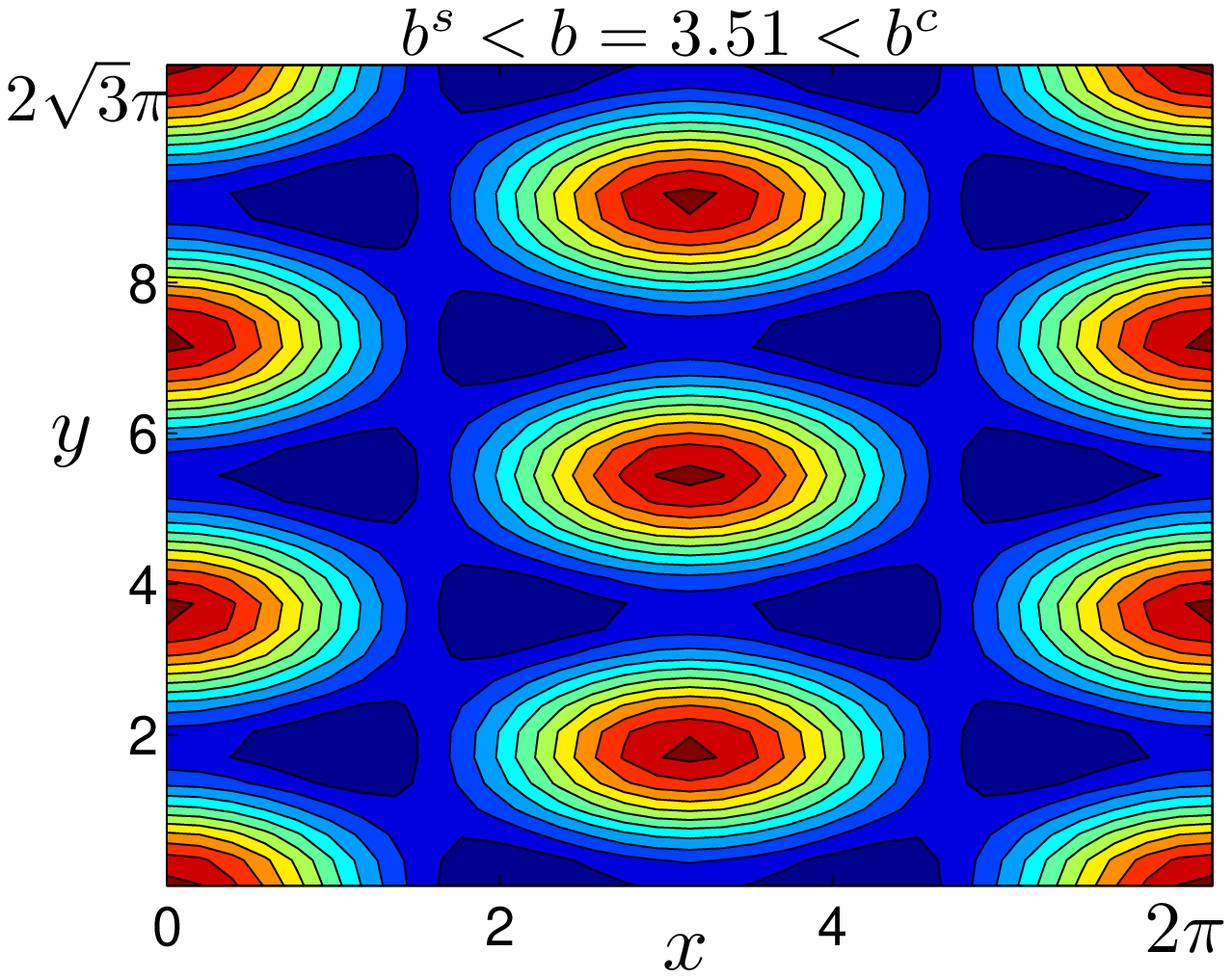} \epsfxsize=5.65cm
\epsfbox{vuoto.eps}
\epsfxsize=5.45cm \epsfbox{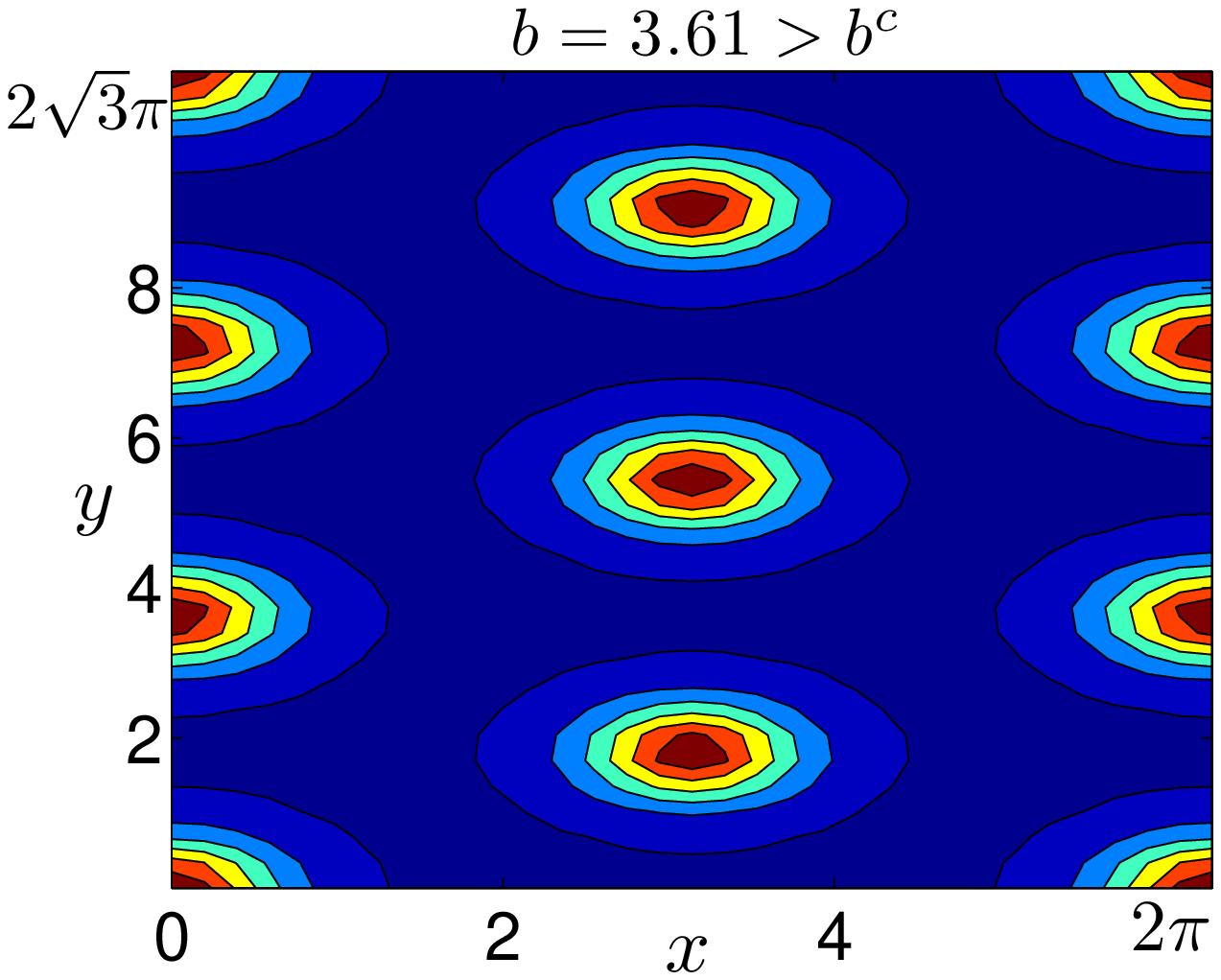}\hskip1.8cm\epsfxsize=5.65cm
\epsfbox{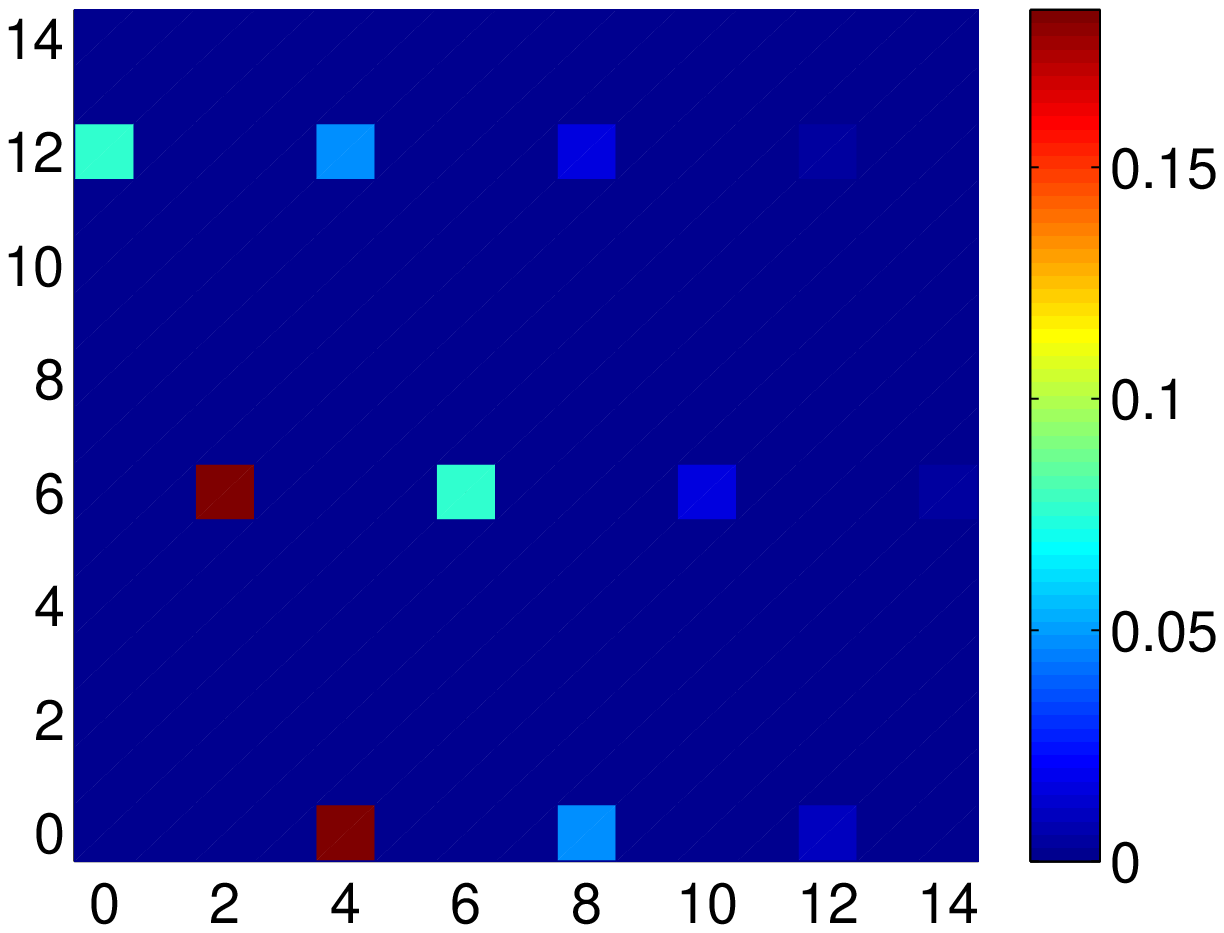}
\caption{Hysteresis cycle. {\it Left.} Numerical solutions. {\it Right.} The spectrum of the solutions.}\label{hyst_esa}
\end{figure}

On the other hand a complete analysis of the existence and stability of the stationary points $H^{\pm}_i$ is too involved
to be carried out in general; instead  we present a detailed numerical study of a typical case.
In Fig.\ref{esagoni} the parameter set has been chosen
such that the only unstable discrete mode is $\bar{k}_c^2=4$, which corresponds, in the domain
$L_x=2\pi$ and $L_y=2\sqrt{3}\pi$, to the
two mode pairs $(2, 6)$ and $(4, 0)$ satisfying the condition \eqref{k2d}. Numerical simulations, performed choosing
as initial condition a random periodic perturbation of the equilibrium, shows that the solution evolves
to the hexagonal pattern on the left of Fig.\ref{esagoni}.
The form of this pattern is captured by the following hexagonal pattern, predicted to be a stable solution via the weakly nonlinear analysis:
\begin{equation}\label{hex1}
{\bf
w}=\varepsilon\bfrho\left({A}_{11}^+\cos(x)\cos\left(\sqrt{3}y\right)+{A}_{21}\cos(2x)\right)+O(\varepsilon^2),
\end{equation}
where $H^+_1\equiv({A}_{11}^+,{A}_{21})$ is a stable equilibrium of the system \eqref{4.32}.
The comparison between the numerical solution and the expected solutions \eqref{hex1} is shown in Fig.\ref{esagoni}.
The WNL analysis predicts very precisely the dominant modes, but the overall accuracy
is spoiled by the presence of sub-harmonics that, as usual for subcritical cases, the WNL analysis underestimates.
%
In Fig.\ref{bif_RH} we show the bifurcation diagram.
One can see that system \eqref{4.32} admits four real roots of the form
$H^{\pm}_j \equiv \left(A^{\pm}_{1j},A_{2j}\right),\,j=1,2$, where $H^{\pm}_1$ are stable and $H^{\pm}_2$ are unstable.
Moreover the system admits as a stable state, also the rolls $R^-$,

In Fig.\ref{bas_RH} we show the basins of attraction of the three stable points $R^-, H^+$ and $H^-$ of the system \eqref{4.32}.
We have run several numerical tests: starting from an initial condition of the form:
\begin{equation}\label{IN_bas}
{\bf w}_1^0=\bfrho\left(A_1^0\cos(\phi_1 x)\cos(\sqrt{3}\phi_1 y)+
A_2^0\cos(2\phi_1 x)\right)
\end{equation}
where $\left(A_1^0, A_2^0\right)$ is in the basin of attraction respectively of $R^-, H^+$ or $H^-$, the solution of the original system \eqref{1.1} evolves to the appropriate stable solution predicted via weakly nonlinear analysis in each case.
The coexistence of more than one stable points for a single value of the bifurcation parameter allows the possibility of hysteresis.
In Fig.\ref{hyst_esa} it is shown this phenomenon: starting with a value of the parameter above $b^c$, the solution jumps
to the stable branch $H^+$ and the hexagonal pattern forms (see Fig.\ref{hyst_esa} with $b=3.67>b^c=3.599$).
Decreasing $b$ slightly below the value $b^c$, the solution persists on the stable branch and the pattern does not disappear
(see Fig.\ref{hyst_esa} with $b^s<b=3.52<b^c$).
With a further decrease below $b^s=3.422$, the solution jumps to the uniform steady state (see Fig.\ref{hyst_esa} with $b=3.42<b^s$)
and  persists in this state even after increasing the parameter $b$ (see Fig.\ref{hyst_esa} with $b^s<b=3.51<b^c$).
To have the pattern formation the parameter must be increased above $b^c$ (see Fig.\ref{hyst_esa} with $b=3.61>b^c$).
On the right of Fig.\ref{hyst_esa}) we show the spectrum of the solution: the dominant modes are those prescribed  by the
weakly nonlinear analysis (i.e. $(2, 6)$ and $(4, 0)$).
The other modes appearing in the Fourier spectrum are the sub-harmonics which are however underestimated by the WNL analysis.

\subsection{Cross-roll instability}

\begin{figure}
\begin{center}
\subfigure[t=0]{\includegraphics[width=6.cm,height=5cm]{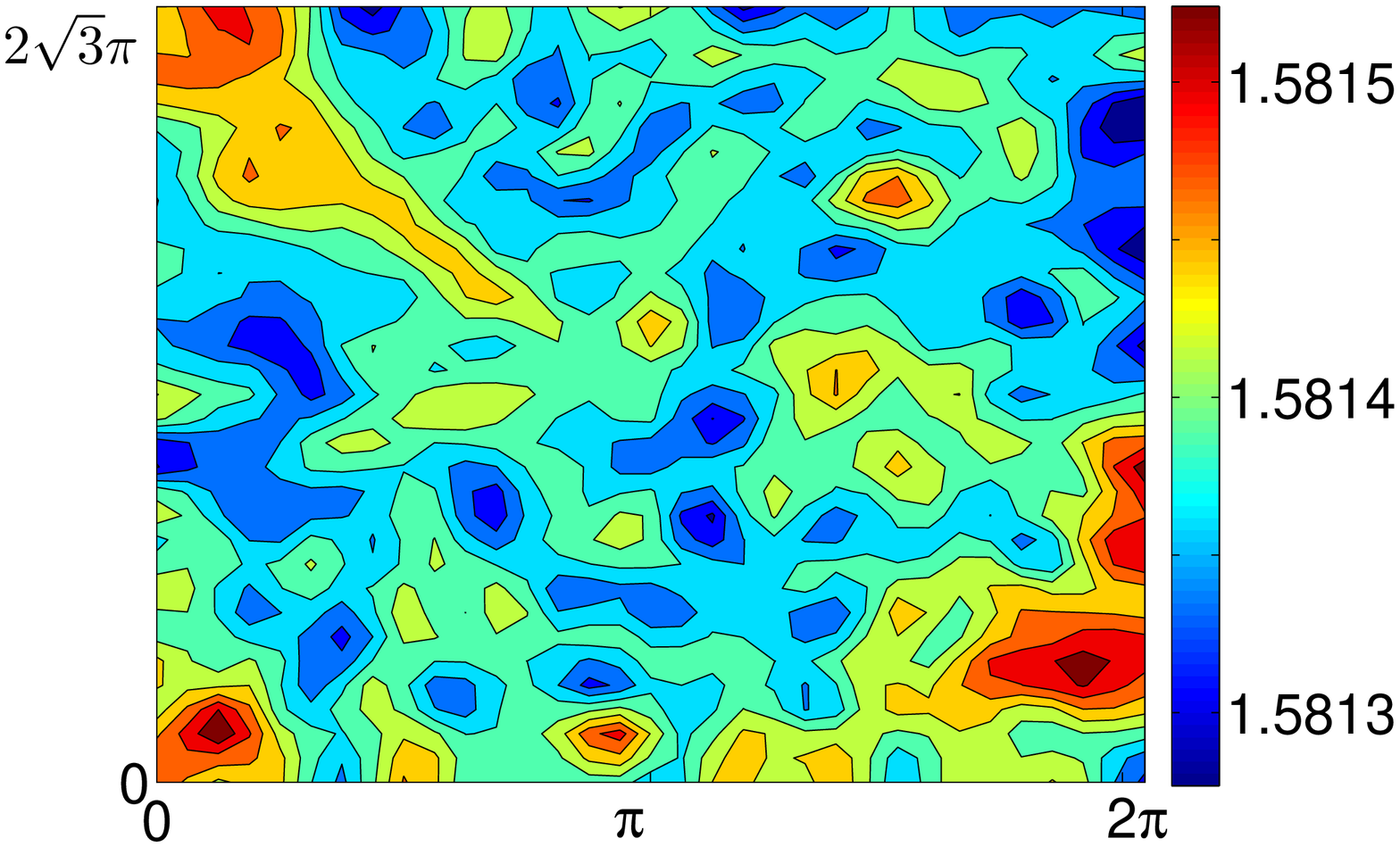}}
\subfigure[t=160000]{\includegraphics[width=6.cm,height=5cm]{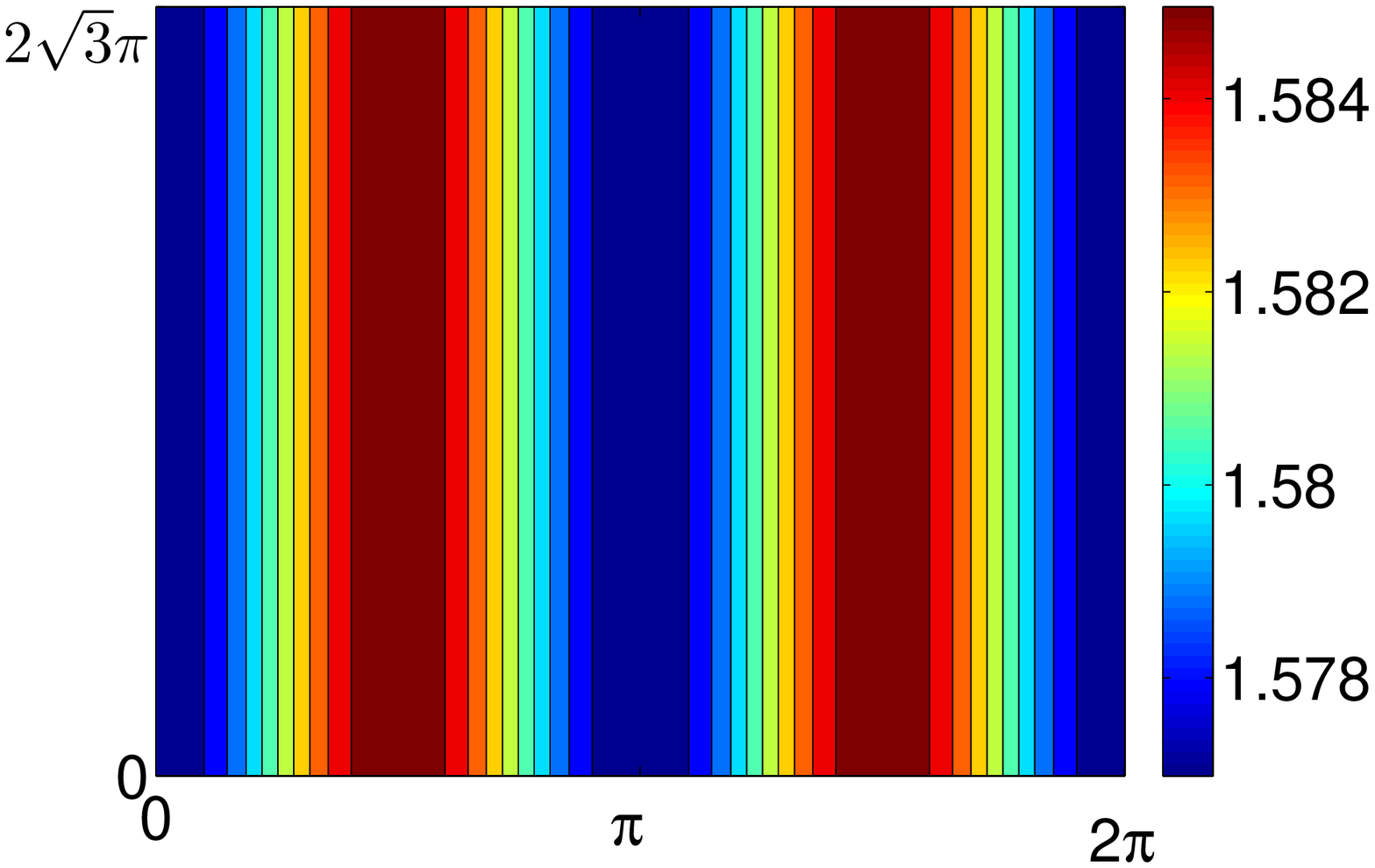}}

\subfigure[t=200000]{\includegraphics[width=6.cm,height=5cm]{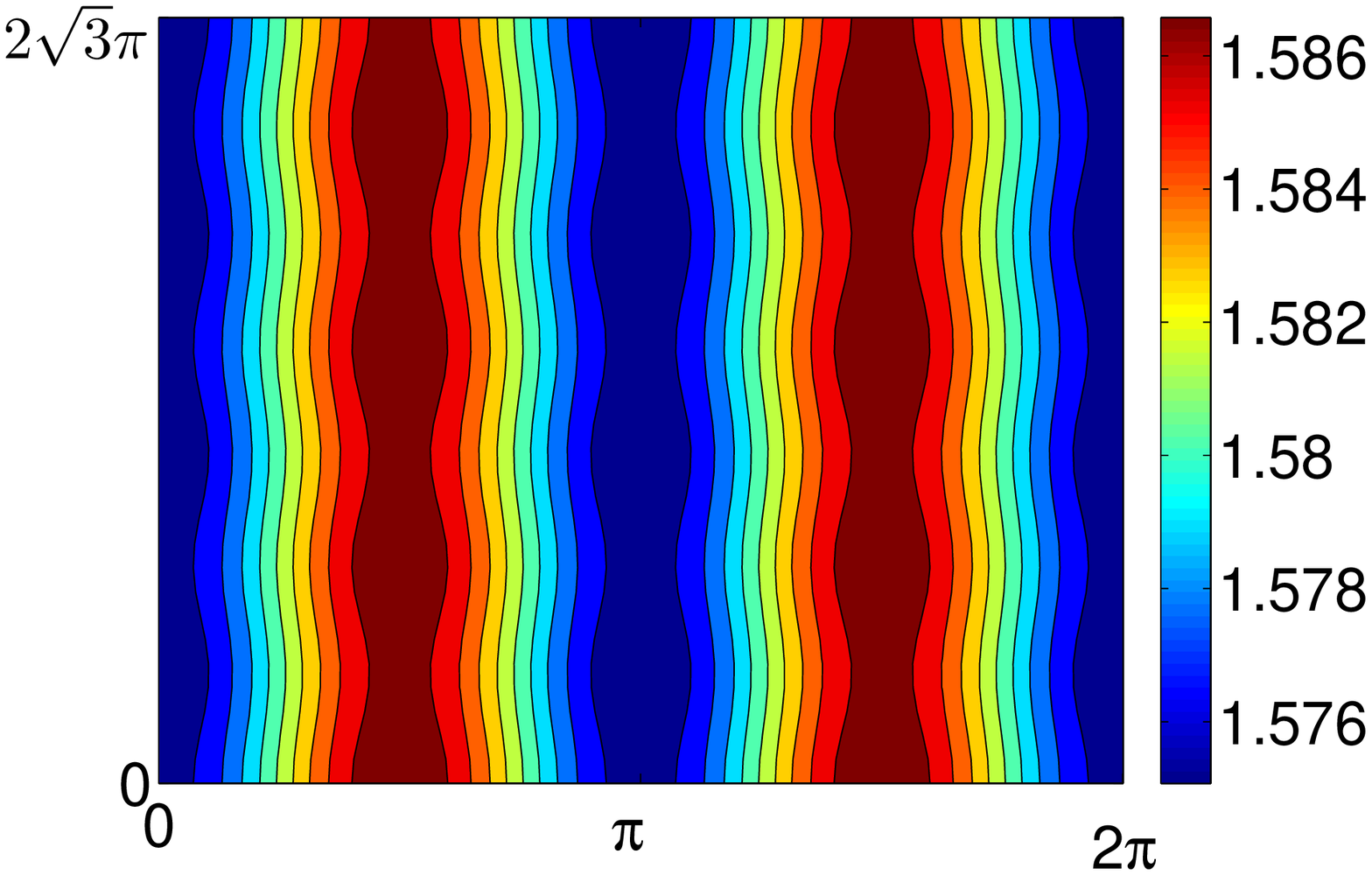}}
\subfigure[t=240000]{\includegraphics[width=6.cm,height=5cm]{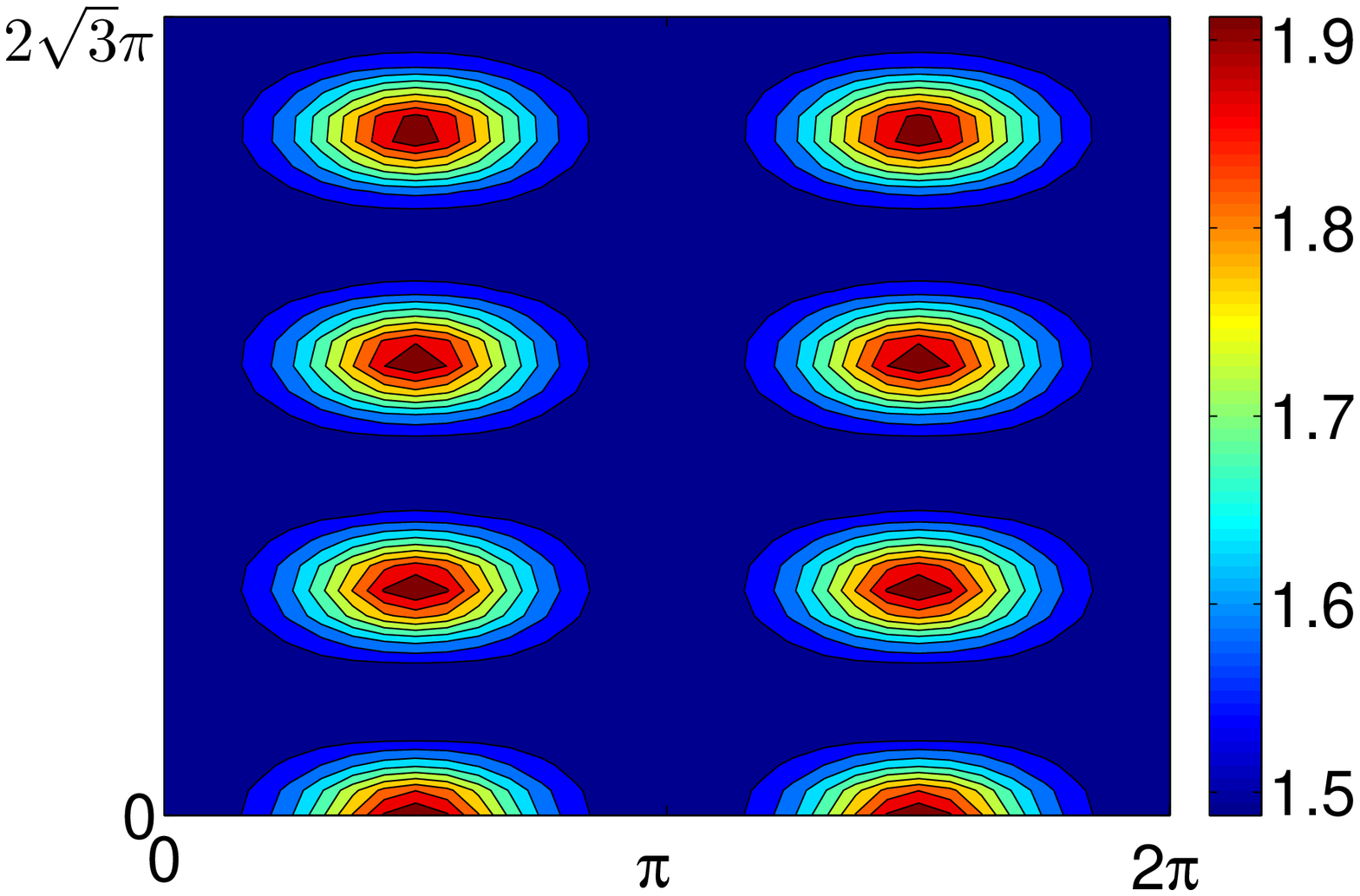}}
\end{center}
\caption{Cross-Roll instability. Starting from a random initial datum the solution evolves to the expected roll pattern \eqref{ex_roll}. However a cross-roll instability leads to a mixed mode pattern.}\label{cross_rollINST}
\end{figure}

We have also observed the phenomenon of the cross-roll instability.
For the same parameter set chosen in Fig.\ref{esagoni} we have performed a numerical simulations almost next to the threshold,
for $\varepsilon=10^{-4}$.
Choosing the initial conditions as a random periodic perturbation of the equilibrium (where the amplitudes of the random modes are small),
the solution evolves, as shown in Fig.\ref{cross_rollINST}(b), to the roll pattern predicted via the weakly nonlinear analysis:
\begin{equation}\label{ex_roll}
{\bf w}=\varepsilon \sqrt{-{\bar{\sigma}_2}/{\bar{\alpha}_2}}\bfrho\cos(2 x)+O(\varepsilon^2).
\end{equation}
Roughly at $t=200000$ the roll pattern loses its stability and the solution evolves to the pattern shown in Fig.\ref{cross_rollINST}(d).

\begin{figure}
\begin{center}
\epsfxsize=6.55cm\epsfbox{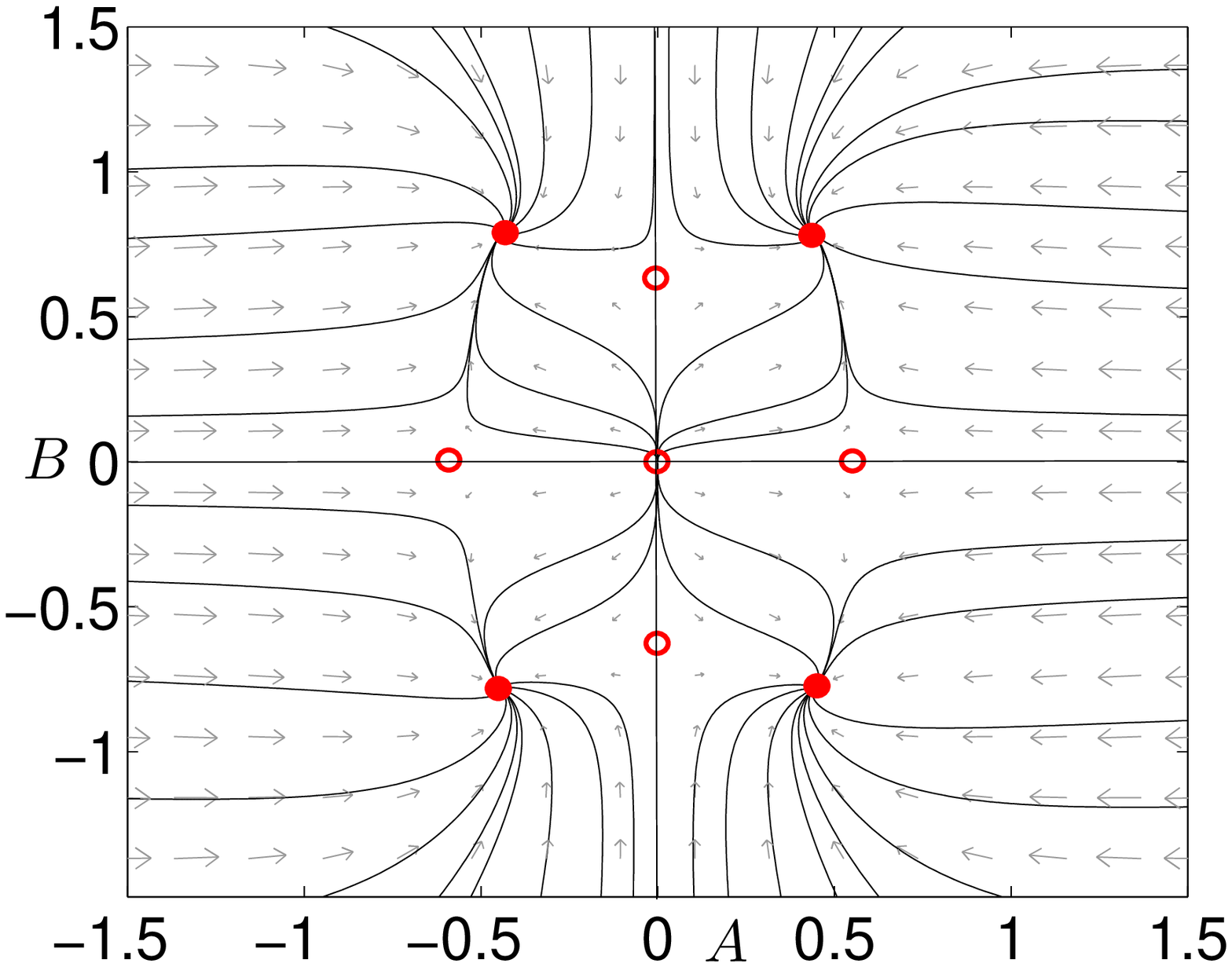}
\end{center}
\caption{The phase plane of system \eqref{comp}. The competitive modes are $\bar{k}_c^2=4$ and $k_m^2=\frac{49}{12}$.}\label{phase}
\end{figure}

A new set of rolls, corresponding to the mode $(0,7)$, grows perpendicularly to the original roll pattern.
In the considered domain $L_x=2\pi$ and $L_y=2\sqrt{3}\pi$, the mode pair $(0,7)$ corresponds, in the sense of \eqref{k2d}, to the eigenvalue $k_m^2=49/12\simeq4.083$, which is nearly equal to the predicted most unstable eigenvalue $\bar{k}_c^2=4$.
We shall interpret the formation of the cross-roll pattern as due to an intermode competition, following the approach presented in \cite{Segel2}.
We perform the weakly nonlinear analysis and write the solution of the linear problem at $O(\varepsilon)$ as:
\begin{equation}\label{comp_roll}
{\bf w}_1=\bfrho\left( A(T_1, T_2)\cos(2 x)+B(T_1, T_2)\cos\left(\frac{7}{2\sqrt{3}} y\right)\right).
\end{equation}
At $O(\varepsilon^3)$ the following ODE model that illustrates the nonlinear behavior of the amplitudes $A$ and $B$ of the two competing modes $\bar{k}_c$ and $k_m$ is obtained:
\begin{subequations}\label{comp}
\begin{eqnarray}\label{comp_a}
\frac{dA}{dT_2}&=&\sigma_1 A-L_1 A^3+\Omega_1 A B^2,\\
\frac{dB}{dT_2}&=&\sigma_2 B-L_2 B^3+\Omega_2 A^2\,
B,\label{comp_b}
\end{eqnarray}
\end{subequations}
Here we skip all the details as the analysis of this system is similar to the case given in Section.\ref{mixedmode}.
For the considered set of parameters, the equilibria $(\pm A, 0)$ and $(0, \pm B)$, each one corresponding
to a roll pattern, are unstable.
The unique stable states are $(\pm A, \pm B)$, as shown into the Fig.\ref{phase}, which corresponds to
the cross-roll pattern.

The above method,  however, does not explain  the fact that the
amplitude of the resulting pattern is $O(1)$ (see Fig.\ref{cross_rollINST}(d)), which would probably require
to consider also the competition with the hexagons modes and the spatial modulation
of the pattern.

\section{Conclusions}

In this paper we have investigated the Turing mechanism induced by nonlinear cross-diffusion
for two coupled reaction-diffusion equations on a two-dimensional spatial domain.

The possibility that the Turing bifurcation occurs via a degenerate eigenvalue makes the study
mathematically involved, but gives rise to a rich variety of patterns which
tessellate the plane and appear as steady state solutions of the reaction
diffusion system.
These are rolls, squares and mixed-mode patterns, among which
there are the supersquares and the hexagons.

We have obtained the amplitude equations providing a mathematical
description of the reaction-diffusion system close to the onset of instability.
The analysis of the amplitude equations has shown the occurrence of
a number of different phenomena, including stable subcritical Turing patterns
or multiple branches of stable solutions  leading to hysteresis.
In particular, hexagonal patterns appear
via a subcritical bifurcation and, from its same primary bifurcation point,
rolls bifurcate supercritically.
Therefore, there is a region of bistability where both rolls and hexagons are stable; however
here rolls appear as a transient state due to a spatially modulated cross-roll instability
that drives the solution toward  a mixed modes pattern.
Our attempt at explaining this instability as due to mode competition has been partially successful.

Future analysis might move from stationary Turing patterns to
traveling patterning waves: when the domain size is large, the
pattern is formed sequentially and traveling wavefronts are the
precursors to patterning. In this case the equations governing
the amplitude of the spatially modulated pattern (one has to consider the slow
modulation in space of the pattern amplitude) will be
the Ginzburg-Landau equation, or systems of Ginzburg-Landau equations in the degenerate case;
we believe that these systems would be of independent and significant mathematical interest.

\section*{Acknowledgements}
The authors thank the referees for the comments and the suggestions that helped improve the paper.
The authors acknowledge the financial support received by INDAM and by the Department of Mathematics, University
of Palermo.

\bibliographystyle{plain}
\bibliography{pattern}

\end{document}